\documentclass[fleqn,usenatbib]{mnras}

\usepackage{newtxtext,newtxmath}

\usepackage[T1]{fontenc}

\DeclareRobustCommand{\VAN}[3]{#2}
\let\VANthebibliography\thebibliography
\def\thebibliography{\DeclareRobustCommand{\VAN}[3]{##3}\VANthebibliography}

\usepackage{graphicx}	
\usepackage{amsmath}	
\usepackage{ltablex}
\usepackage{longtable}

\usepackage{soul}
\renewcommand\hl

\title[Internal properties of UCDs]{Testing the tidal stripping scenario of ultra-compact dwarf galaxy formation by using internal properties}

\author[R. J. Mayes et al.]{Rebecca J. Mayes,$^{1}$\thanks{E-mail: r.mayes@uq.net.au}
Michael. J. Drinkwater,$^{1}$
Joel Pfeffer,$^{2}$ Holger Baumgardt,$^{1}$  
\newauthor Chengze Liu,$^{3}$ Laura Ferrarese,$^{4}$ Patrick C{\^o}t{\'e},$^{4}$ and Eric W. Peng$^{5,6}$
\\
$^{1}$School of Mathematics and Physics, University of Queensland, Brisbane, QLD 4072, Australia\\
$^{2}$International Centre for Radio Astronomy Research (ICRAR), M468, University of Western Australia, 35 Stirling Hwy, Crawley, WA 6009, Australia\\
$^{3}$Department of Astronomy, School of Physics and Astronomy, and Shanghai Key Laboratory for Particle Physics and Cosmology, \\
Shanghai Jiao Tong University, Shanghai 200240, P. R. China\\
$^{4}$Herzberg Astronomy and Astrophysics Research Centre, National Research Council of Canada, Victoria, BC V9E 2E7, Canada\\
$^{5}$Department of Astronomy, Peking University, Beijing China 100871\\
$^{6}$ Kavli Institute for Astronomy and Astrophysics, Peking University, Beijing China 100871\\
}

\date{Accepted XXX. Received YYY; in original form ZZZ}

\pubyear{2015}

\begin{document}
\label{firstpage}
\pagerange{\pageref{firstpage}--\pageref{lastpage}}
\maketitle

\begin{abstract}
We use the hydrodynamical EAGLE simulation to  test if ultra-compact dwarf galaxies (UCDs) can form by tidal stripping by predicting the ages and metallicities of tidally stripped galaxy nuclei in massive galaxy clusters, and compare these results to compiled observations of age and metallicities of observed UCDs. We further calculate the colours of our sample of simulated stripped nuclei using SSP models and compare these colours to observations of UCDs in the Virgo cluster. We find that the ages of observed UCDs are consistent with simulated stripped nuclei, with both groups of objects having a mean age > 9 Gyr. Both stripped nuclei and UCDs follow a similar mass-metallicity relationship, and the metallicities of observed UCDs are consistent with those of simulated stripped nuclei for objects with M > 10\textsuperscript{7}~\(\textup{M}_\odot\). The colours of observed UCDs are also consistent with our simulated stripped nuclei, for objects with M > 10\textsuperscript{7}~\(\textup{M}_\odot\), with more massive objects being redder. We find that the colours of stripped nuclei exhibit a bimodal red and blue distribution that can be explained by the dependency of colour on age and metallicity, and by the mass-colour relation. We additionally find that our low mass stripped nuclei sample is consistent with the colour of blue globular clusters. We conclude that the internal properties of simulated nuclei support the tidal stripping model of UCD formation.
\end{abstract}




\section{Introduction}

Ultra-compact dwarf galaxies (UCDs) are a class of stellar system first discovered in the Fornax cluster \citep{Hilker1999, Drinkwater2000} and have since been discovered in other clusters and galaxy groups such as Virgo \citep{has2005, liu2020generation}, Abell 1689 \citep{Mieske2005}, Centaurus \citep{Mieske2007a}, Hydra \citep{Wehner_2007}, Abell S0740 \citep{Blakeslee_2008}, the NGC 1023 group \citep{Mieske2007b}, the Dorado group \citep{Evstig2007a},  the NGC 5044 group \citep{Faifer_2017},
the NGC 3613 group \citep{De_B_rtoli_2020} and the
NGC 1132 fossil group \citep{Madrid_2011}. They have also been discovered around isolated galaxies \citep{Hau200}. 

The exact definition for a UCD is somewhat arbitrary, but they generally have properties intermediate between globular clusters and dwarf galaxies, with absolute magnitudes -14.0~mag < M\textsubscript{v} < -10~mag \citep{Voggel2016} and half-light radii 7~pc < r\textsubscript{h} < 100~pc \citep{Mieske2008}. They have central velocity dispersions similar to dwarf galaxies, approximately 20 < $\sigma$\textsubscript{0} < 50~km s\textsuperscript{-1}, giving dynamical masses of approximately 2~$\times$~10\textsuperscript{6} to 10\textsuperscript{8}~\(\textup{M}_\odot\)
\citep{has2005, Hilker2008, Mieske2008, Mieske2013}. Typically they have old stellar populations of at least 8 Gyr \citep{Chilingarian2011, Norris2015}.

There are a few different theories about how UCDs form, with most scenarios suggesting either a galaxy or star cluster origin. In the first scenario, UCDs are the nuclei of dwarf galaxies stripped by tidal interactions with other galaxies \citep{Bassino1994, Bekki2001, Bekki2003, Drinkwater2003, Goerdt2008, Pfeffer2013, Pfeffer2014}. In the latter scenario UCDs represent the high-mass end of the globular cluster mass function observed around galaxies with extensive globular cluster systems \citep{Mieske2002, Mieske2012}. Additionally, since UCDs are larger than typical globular clusters, they may have formed from the merger of many globular clusters in star cluster complexes \citep{Kroupa1998, Fellhauer2002, brun2011, brun2012}.
There is, indeed, a growing amount of evidence that UCDs are not formed from a single scenario but instead from a combination of the scenarios outlined above \citep{Mieske2006, brodie2011, Chilingarian2011, DaRocha2011,Norris2011,Pfeffer2014, Liu2015, Pfeffer2016, liu2020generation}. 

Stripped nuclei are likely to make up at least some percentage of the UCD population, due to the resemblances UCDs bear to compact nuclei. UCDs have similar internal velocity dispersions \citep{Drinkwater2003} and overlap the luminosity distribution of nuclei \citep{Drinkwater2004} and follow a size-luminosity distribution, being approximately 2.2 times larger than galaxy nuclei at the same luminosity \citep{Evstig2008A}, which may be a consequence of tidal stripping \citep{Pfeffer2013}. UCDs have high dynamical mass-to-light ratios \citep{Forbes_2014, Janz_2015}. They are observed to have colour-magnitude and mass-metallicity relations \citep[e.g.][]{Cote2006, brodie2011,Zhang2018}), to lack colour gradients \citep{Liu2015} and show stellar population similarities to nuclei \citep{Paudel2010, Janz_2015}. 

Many UCDs have been observed to have asymmetric/tidal features \citep{Jennings2015, Mihos_2015,Voggel2016,Schweizer_2018,liu2020generation}, while others show evidence of stellar envelopes or of being transition objects from dwarf galaxies to UCDs \citep{Drinkwater2003, has2005, Penny_2014, Liu2015,liu2020generation}. Some UCDs appear to have globular clusters surrounding them \citep{Voggel2016}. Recently a number of UCDs have been discovered to host supermassive black holes \citep{Seth2014, Ahn2017, Ahn2018, Afanasiev_2018} which make up a large percentage of the UCD mass, which is a strong indication of a tidal stripping origin.

N-body simulations have demonstrated the viability of dwarf galaxies transforming into UCDs \citep{Bekki2001, Bekki2003, Pfeffer2013}, and that stripped nuclei can explain UCD distributions and make up a significant percentage of the high mass UCD population \citep{Pfeffer2014, Pfeffer2016, Mayes2020}.

In \citet{Mayes2020} (hereafter Paper I) we presented the first model for UCD formation based on tidal stripping of dwarf galaxies in a cosmological hydrodynamical simulation. We then determined the numbers and radial distributions of simulated stripped nuclei in Virgo-sized galaxy clusters, and found that they were consistent with the numbers and radial distributions of UCDs observed in the Virgo cluster.

In this paper we further test the model by using the baryonic component of the EAGLE simulation to determine the ages, colours and metallicities of our simulated stripped nuclei from Paper I. We then compare this sample with the internal properties of observed UCDs and globular clusters. Our aims are to determine whether the ages, metallicities and colours of simulated stripped nuclei can match those of observed UCDs, whether stripped nuclei exhibit bimodal colour and metallicity distributions such as those observed in globular cluster systems, and whether low mass stripped nuclei are consistent with making up some percentage of the globular cluster population. Throughout the paper we refer to the objects formed in the simulation by the tidal stripping of galaxies as stripped nuclei, because these objects may resemble both globular clusters and UCDs, and UCDs may form by more than one formation channel.

This paper is organized as follows. Section 2 briefly summarizes our stripped nuclei formation method from Paper I, and how we determine further properties of that sample. The results of our research are presented in Section 3. We discuss the implications of our work for observed UCDs and GCs in Section 4, and a summary of our results is given in Section 5.
\section{Methods and Observations}

Here we summarize the stripped nuclei formation model of Paper I, and describe how we determined the ages, metallicities and colours of the simulated stripped nuclei.

\subsection{Creating a sample of simulated stripped nuclei}

We used the hydrodynamical EAGLE simulations to examine stripped nuclei formation. The largest of the simulations, RefL0100N1504 has a box side length of 100 comoving Mpc, made up of 6.8 billion particles. 

We used three sets of data from the EAGLE simulation in this paper: an online database \citep{McAlpine2016}, which contains properties of large scale objects in the simulation like galaxies and clusters, the raw particle data which was used to determine the locations of the simulated stripped nuclei, and a database we created that links the online database and the particle data.

UCDs are too compact to be fully resolved in the EAGLE simulations, but the baryonic particle mass in the simulations, 1.81~$\times$~10\textsuperscript{6}~\(\textup{M}_\odot\), is similar to the mass range of observed UCDs (2~$\times$~10\textsuperscript{6}~\(\textup{M}_\odot\) to 10\textsuperscript{8}~\(\textup{M}_\odot\)). Therefore, we defined the central, most bound star particle (MBP) of a galaxy before a merger as the nucleus of the galaxy and tracked this particle after the merger as a stripped nucleus.

To create a sample of candidate galaxies that could be disrupted to become stripped nuclei, we carried out the following steps. 

\begin{enumerate}
  \item Massive galaxies in clusters at z = 0 were first identified. These will have potentially disrupted smaller galaxies in the past. In Paper I, we chose galaxies with a stellar-mass greater than 10\textsuperscript{7}~\(\textup{M}_\odot\), which approaches the lower limit at which galaxies can be defined in the EAGLE simulation. In this paper we raised our lower limit to galaxies with stellar mass >  10\textsuperscript{7.5}~\(\textup{M}_\odot\) because lower mass galaxies in EAGLE have less reliable properties \citep{Schaye2015}. This will mean we undersample the number of lower mass stripped nuclei, however it increases our resolution as below M =  10\textsuperscript{7.5}~\(\textup{M}_\odot\) galaxies consist of less than 30 particles giving increased scatter in metallicity for a given stellar mass. 
  \item We further created a sample of mergers involving any progenitor galaxies with stellar mass >  10\textsuperscript{7.5}~\(\textup{M}_\odot\). \citet{Sanchez2019} found that at a stellar mass of 10\textsuperscript{7.5}~\(\textup{M}_\odot\) approximately 45 per cent of galaxies are nucleated and have a mean nuclear star cluster to galaxy mass ratio of 1.3 per cent. Therefore, our simulated stripped nuclei sample will be unreliable below a mass of $\approx$ 4~$\times$~10\textsuperscript{5}~\(\textup{M}_\odot\). Several objects above this mass are also lost due to scatter in the relations. 
  \item The particle data was then used to determine properties of the stripped nuclei.
\end{enumerate}

A stripped nucleus is formed in a merger if the following conditions are satisfied:

\begin{enumerate}
  \item The mass ratio between the merged galaxies is < 1/4. In a minor merger, only the less massive galaxy will be disrupted and could therefore leave behind a stripped nucleus \citep{Qu2017}.  
  \item The stripped nucleus has been orbiting the more massive galaxy after the merger for a time shorter than its dynamical friction timescale. We calculated the dynamical friction timescale with equation 7-26 from \citet{Binney1987}, modified with an eccentricity function as defined in Appendix B of \citet{Lacey1993}. 
  Stripped nuclei with dynamical friction timescales shorter than the orbital time will have merged with the central galaxy before the final (z = 0) snapshot.
\end{enumerate}

The number of nucleated galaxies was estimated from Figure 2 from \citet{Sanchez2019}. When considering the nucleation fraction, we primarily work with fractions of stripped nuclei rather than choosing a random sample of stripped nuclei in each mass range. Nuclei mass estimates were based on Figure 9. from \citet{Sanchez2019}, which plots the ratio of nuclear star cluster to galaxy mass as a function of galaxy stellar mass. The stripped nuclei are each assigned a mass randomly chosen from a log-normal mass function for the nucleus-to-galaxy mass ratio with a mean chosen by applying linear fits to the upper and lower mass ranges of Figure 9 in \citet{Sanchez2019} and a log-normal standard deviation of 0.4 dex.

Because of the mass-metallicity and mass-colour relationships of galaxy nuclei our choice of nuclear mass to galaxy stellar mass ratio has an impact on the colours and metallicities of simulated stripped nuclei. In \citet{Sanchez2019} the mass of a nucleus varies depending on galaxy mass, reaching a minimum of approximately 0.36 per cent for galaxies with stellar mass 3~$\times$~10\textsuperscript{9}~\(\textup{M}_\odot\) and then increasing for more and less massive galaxies. \citet{Pfeffer2014} and \citet{Pfeffer2016} adopted a broad nuclear mass to galaxy mass ratio of 0.3 per cent \citep{Cote2006} or 0.1 \citep{Georgiev_2016} per cent. This means that at, for example, a progenitor galaxy mass of 1~$\times$~10\textsuperscript{8}~\(\textup{M}_\odot\), \citet{Sanchez2019} observes a mean nucleus to galaxy stellar mass ratio of 1 per cent, $\approx$ 3 times larger than \citet{Cote2006} and 10 times larger than \citet{Georgiev_2016}. The studies of \citep{Cote2006} and \citep{Georgiev_2016} are limited however, by a lack of lower mass galaxies, so the relation for higher mass galaxies is extrapolated to the lower mass ranges. A nucleus to galaxy stellar mass ratio closer to these two studies would reduce our predicted number of blue, low metallicity objects produced by low mass galaxies.

\subsection{Determination of age, metallicity and colour for the simulated nuclei}
\label{section:agesmets}


The appeal of the EAGLE simulation is that the internal properties of galaxies in the simulation can be determined from the particle data. Each star particle has a defined smoothed metallicity, iron abundance, hydrogen abundance and stellar formation time. We determined these properties of the stripped nuclei by averaging the values for the 5 most central particles of the original galaxy before the merger. The choice of the 5 innermost particles was a compromise between the increase in scatter from the use of a single particle, and the reduced sampling of a galaxies centre from the use of multiple particles.

A challenge with determining the metallicity of the particles within the EAGLE simulation is that low mass EAGLE galaxies are systematically more metal rich than observed galaxies \citep{Schaye2015}. We calculated a mass-dependent metallicity correction to make our mean metallicities consistent with observations following \citet{Kirby_2013} (see Appendix \ref{appendix:metcorrec} for details.)

We also note that our correction is calculated using low redshift galaxies, while the galaxies that make up stripped nuclei predominantly form and merge at high redshift. If high redshift galaxies in the EAGLE simulation are more akin to observed galaxies than low redshift ones, or require a different correction, our stripped nuclei metallicities may be incorrect. Additionally, at a given stellar mass, galaxies in clusters can be slightly more metal rich than those in low mass environments \citep[e.g.][]{Gallazzi2021}. Virgo galaxies from \citet{Chilingarian2009} are found to be approximately 0.3 dex more metal rich than those from \citet{Kirby_2013} at the same stellar mass, which could cause us to overestimate the correction required. However, as stripped nuclei primarily form and merge in the early stages of the universe, galaxies in low mass environments may be a more accurate comparison.

In addition to the stellar formation age that is recorded by the EAGLE particles, we also consider the merger time age of the stripped nuclei, which is the time at which the original progenitor galaxy was disrupted and the nucleus became stripped. This is done because of the possibility that when a merger occurs the stripped nucleus may experience a starburst that the EAGLE simulation does not account for. We call this the merger time (MT) age, and the age recorded by the EAGLE particles the stellar formation time (SFT) age.

We calculated the colours of the stripped nuclei sample from the ages and metallicities using the Flexible Stellar Population Synthesis (FSPS) model  \citep{Conroy2009,Conroy2010} assuming the default spectral library and isochrones, and a Chabrier stellar initial mass function (IMF) \citep{Chabrier2003}.
Luminosities were calculated in the CFHT MegaCam filters (u*, g, r, i and z) by linearly interpolating from the grid in ages and metallicities.
Colours can then be calculated; in this study we consider the colours u*-g, u*-i, u*-g, g-i, g-z and i-z, that are used in \citet{liu2020generation}. We note that the u* magnitudes are difficult to calculate accurately \citep{Choi2019} so we use the results from it with caution. Uncertainties in the metallicity will affect the colour calculation, however, as shown in Fig.~\ref{fig:gimetage}, this is smaller than the impact on metallicity.

\subsection{Observational Data} 
\label{obsdata}
To compare our simulated stripped nuclei properties with observed UCD properties we first created two samples of UCD properties from the literature.

The metallicities and ages of UCDs were compiled from \citet{Mieske_2008,Paudel2010,Francis2012,Chilingarian2011} and \citet{Forbes_2020}. Where studies contained duplicate objects priority was given first to objects with properties measured by spectral fitting and then to more recent papers. A number of objects were found to have unreasonably large ages > 13.7 Gyrs. These were assigned a random age between 11 Gyrs and 13.7 Gyr due to the difficulty in accurately measuring extremely old objects. Objects in the studies of \citet{Chilingarian2011} and \citet{Francis2012} do not have given masses, so for those for which we could not obtain masses from \citet{Mieske_2008}, we performed a mass calculation using M\textsubscript{v} and M/L ratio as measured by \citet{Chilingarian2011} and a more crude calculation with the g band magnitude and an assumed M/L ratio of 2.15 \(\textup{(M/L)}_\odot\) \citep{Voggel_2019}, for \citet{Francis2012}. Comparisons of our mass calculation with duplicate objects in \citet{Mieske_2008} found that the masses we calculated for \citet{Chilingarian2011} UCDs are fairly accurate, however the masses we calculate using the data from  \citet{Francis2012} are too low by a factor of 2-3, as a result of the limited data for the calculation. The offset was found to be more pronounced for the more massive objects, which may indicate an underestimation of the mass-to-light ratio. 

Our compiled sample of UCD ages and metallicities can be found in Appendix Table~\ref{tab:agesmets}.

Our colour analysis was made using a new sample of 243 UCD candidates in the Virgo cluster \citep{liu2020generation}. The completeness level of the sample is over 90 per cent and selects objects brighter than g~$\leqslant$~21.5~mag (or more massive than M~=~1.6~$\times$~10\textsuperscript{6}~\(\textup{M}_\odot\)). Objects are selected with a radius of > 11 pc, which will reduce the amount of lower mass objects relative to our study, as the mass-size relation of UCDs reaches 10~pc at a magnitude of  M\textsubscript{v}~$\approx$~-11~mag, or mass M~$\approx$~10\textsuperscript{7}~\(\textup{M}_\odot\) \citep{Norris2011}.

\section{Results}

Because observations of UCDs are more complete at higher masses we carry out our comparisons of stripped nuclei and UCDs with two separate samples. The low mass sample includes all UCDs and stripped nuclei with M > 2~$\times$~10\textsuperscript{6}~\(\textup{M}_\odot\) and may be an incomplete sample of UCDs. The high mass sample is defined as objects with M > 1~$\times$~10\textsuperscript{7}~\(\textup{M}_\odot\), above which the \citet{liu2020generation} sample is believed to be mostly complete.

\subsection{Ages and merger times}
\label{section:ages}

\begin{figure}
	\includegraphics[width=\columnwidth]{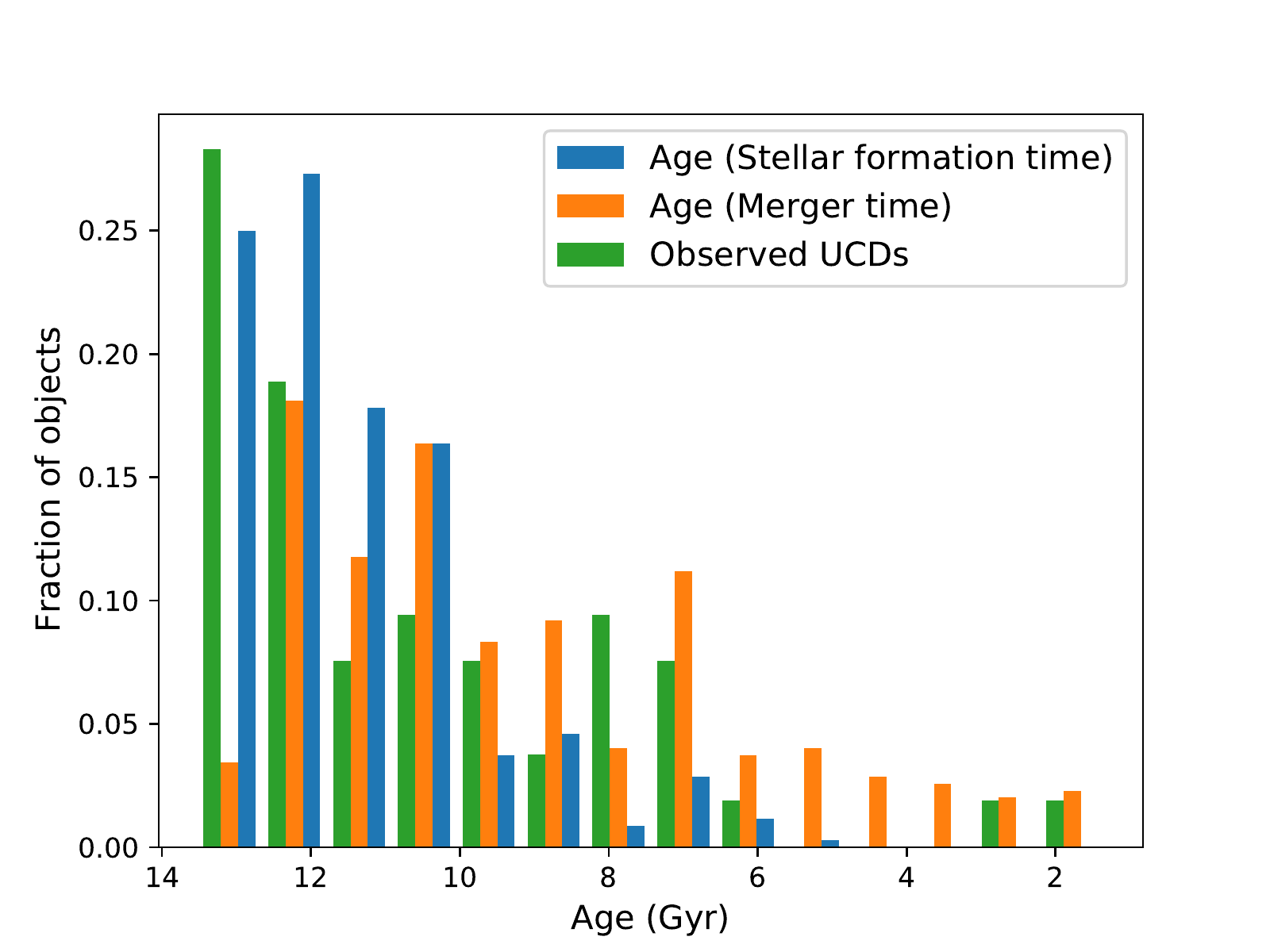}
    \caption{Comparative age distributions for the simulated stripped nuclei stellar formation time, merger time, and the ages of observed UCDs from Table~\ref{tab:agesmets}. In this plot, UCDs with age measurements greater than the age of the universe are assigned a random age between 11 and 13.7 Gyr.}
    \label{fig:projages}
\end{figure}

Fig.~\ref{fig:projages} depicts the stellar formation and merger ages of simulated stripped nuclei, as compared with observed UCDs from Table~\ref{tab:agesmets}. Stellar formation time is the age at which the particles that make up a stripped nucleus formed in the simulation and is the most comparable property to observations. The merger time is the time at which a progenitor galaxy was disrupted to form a  stripped nucleus. As noted above (Sec.~\ref{section:agesmets}) we use the merger time as a proxy for merger-induced star formation not resolved in the EAGLE data. The plot shows that stripped nuclei in the simulations form very early, with none being younger than 5 Gyr, and over 80 per cent being older than 10 Gyr, although mergers continue to occur up to 1 Gyr. The mean star formation age was 11.47 $\pm$ 0.09 Gyr, while the mean merger time was 9.0 $\pm$ 0.2 Gyr, indicating that stripped nuclei primarily form and merge in the early stages of the universe. 
Observed UCDs are also typically old, with the mean age being 10.6 $\pm$ 0.4 Gyr. This calculation is dependent on our assigning UCDs with estimated ages greater than 13.7 Gyr (corresponding to the age of the universe) a random age between 11 and 13.7 Gyr. If these are instead set at 13.7 Gyr, the mean age becomes 11.1 $\pm$ 0.4 Gyr, consistent with the ages of simulated stripped nuclei. 

However, directly comparing the age spectrum of simulated stripped nuclei and observed UCDs is difficult because measuring precise ages for objects older than 9 Gyr is extremely challenging \citep[e.g.][]{Conroy_2013, Spengler2017}, due to stellar population uncertainties.
We found that the ages of UCDs in the literature vary strongly from study to study. Even within single studies using different methods of measuring ages \citep[e.g.][]{Francis2012}, ages can be inconsistent, with age measurements made by lick indices and spectral fitting differing by up to 5 Gyrs for a single UCD, even considering errors given. Therefore it is difficult to make specific comparisons between the simulated stripped nuclei and observed UCD ages. There do appear to exist several younger UCDs which are inconsistent with the stellar formation time, although these represent only a small number of objects. These young objects could be explained if some stripped nuclei experience a significant starburst at merger that is not accurately measured in EAGLE due to the low particle resolution.



\subsection{Stripped nuclei with masses \texorpdfstring{M $>$ 2~$\times$~10\textsuperscript{6}~\(\textup{M}_\odot\)}{Lg}}

\subsubsection{Metallicities}
\label{section:midmets}
\begin{figure}
	\includegraphics[width=\columnwidth]{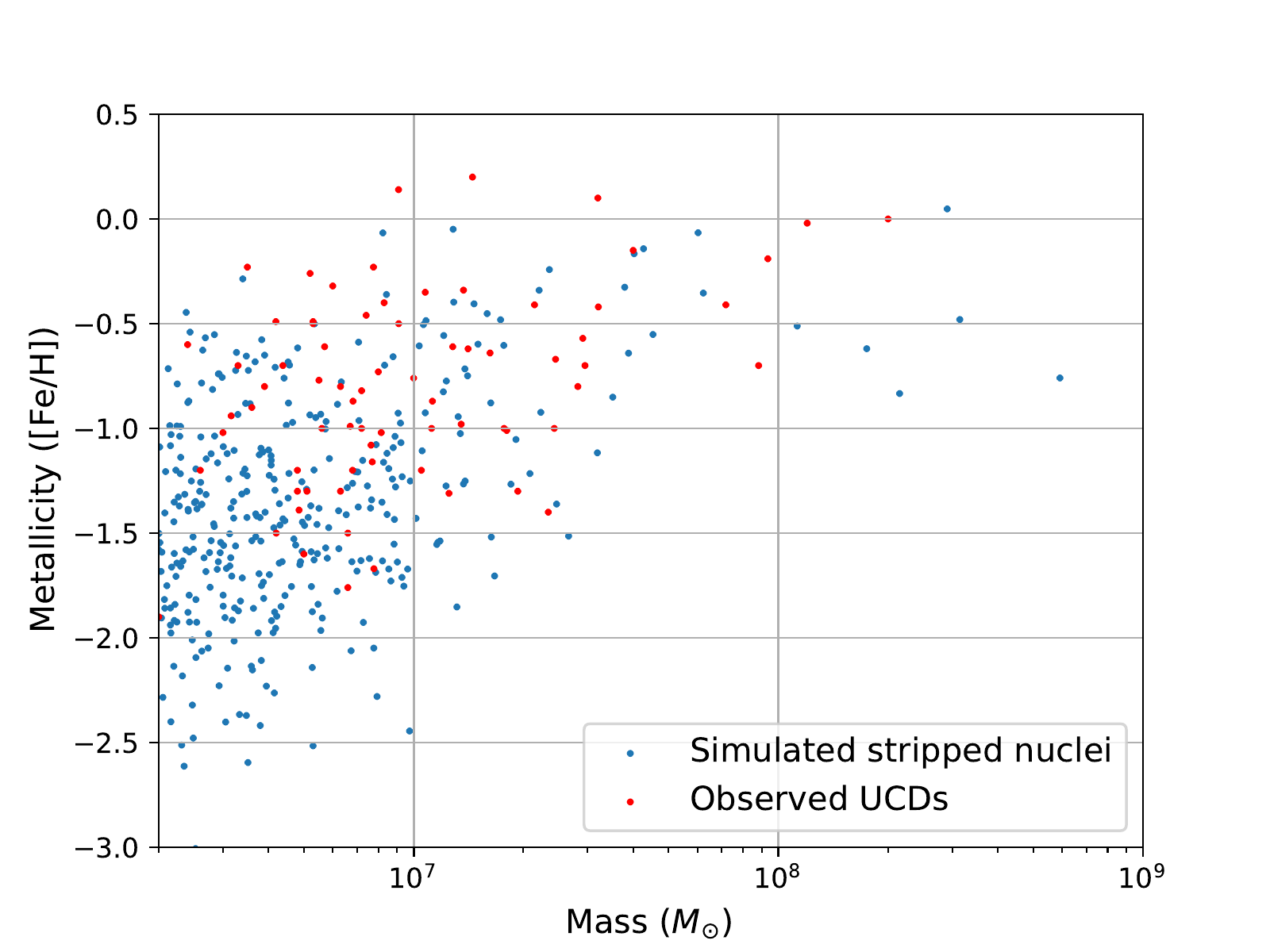}
    \caption{Metallicity ([Fe/H]) against stellar mass for simulated stripped nuclei and observed UCDs from Table.~\ref{tab:agesmets}}
    \label{fig:femassplot}
\end{figure}

Fig.~\ref{fig:femassplot} plots mass against metallicity for our sample of simulated stripped nuclei and observed UCDs. There is significant overlap between the two populations. The metallicities of all the UCDs are consistent with those measured for stripped nuclei within the uncertainties. Both observed UCDs and simulated stripped nuclei appear to follow a mass-metallicity relationship, with metallicity increasing with mass, that is especially strong for objects with mass M $>$ 1~$\times$~10\textsuperscript{7}~\(\textup{M}_\odot\). 

This plot also shows that we predict a number of low metallicity stripped nuclei that are not currently observed. This may be a result of the limited number of metallicity measurements that have been made for the population of low mass UCDs, and the uncertainties in EAGLE metallicities for low-mass galaxies. An alternative explanation is that since the divide between globular clusters and UCDs is fairly arbitrary and our stripped nuclei mass estimates are approximations based on the progenitor galaxy stellar mass, these low mass stripped nuclei would be observed as metal poor (blue) globular clusters.

The majority of simulated stripped nuclei are extremely metal poor, although there are a small population of more metal rich objects. 

Considering the metallicities of only high mass simulated stripped nuclei with M > 1~$\times$~10\textsuperscript{7}~\(\textup{M}_\odot\) a Kolmogorov-Smirnov (KS) test 
does not find that the two distributions are inconsistent (p = 0.22)
The mean [Fe/H] for the simulated objects with M~>~1~$\times$~10\textsuperscript{7}~\(\textup{M}_\odot\) is -0.86 $\pm$ 0.07, while the mean [Fe/H] of observed UCDs with M~>~1~$\times$~10\textsuperscript{7}~\(\textup{M}_\odot\) is -0.63 $\pm$ 0.08. Note that the slight underestimation of masses of the 13 UCDs from the \citet{Francis2012} sample may mean some UCDs may have been excluded from this sample, and including them would bring the mean metallicities of simulated nuclei and observed UCDs closer together. 
 
Fitting a one-dimensional Gaussian mixture model to the stripped nuclei metallicity data found no evidence for a bimodal metallicity distribution, like that possibly exhibited by globular clusters \citep[e.g.][]{Forbes1997, Cote_1998, Brodie2006, Peng_2006, Forbes_2011}.

\subsubsection{Colours}
\label{section:midcolours}

\begin{figure}
	\includegraphics[width=\columnwidth]{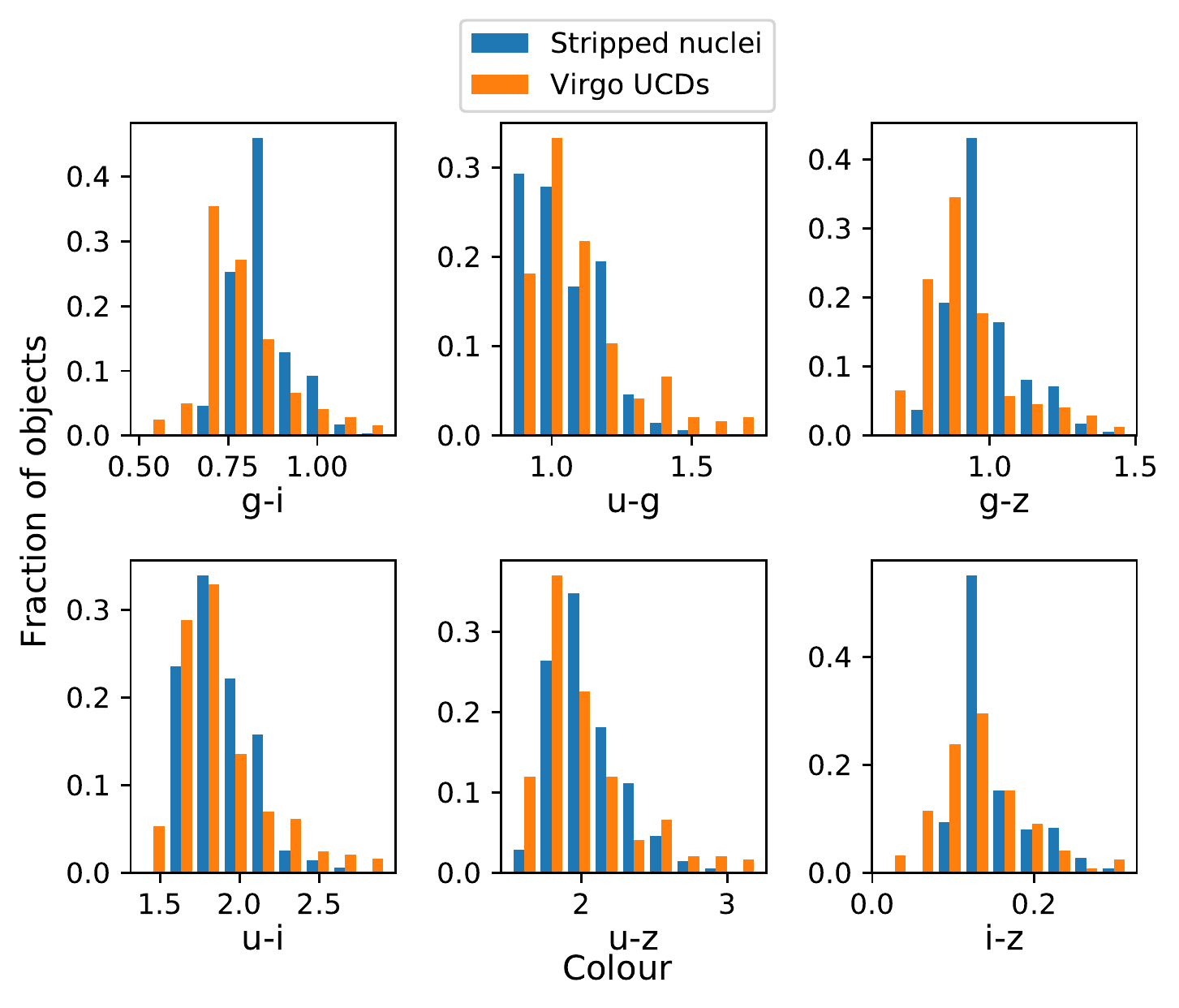}
    \caption{Colours of M > 2~$\times$~10\textsuperscript{6}~\(\textup{M}_\odot\) simulated stripped nuclei from our study and observed UCDs from \citep{liu2020generation}.}
    \label{fig:cols}
\end{figure}

\begin{figure}
	\includegraphics[width=\columnwidth]{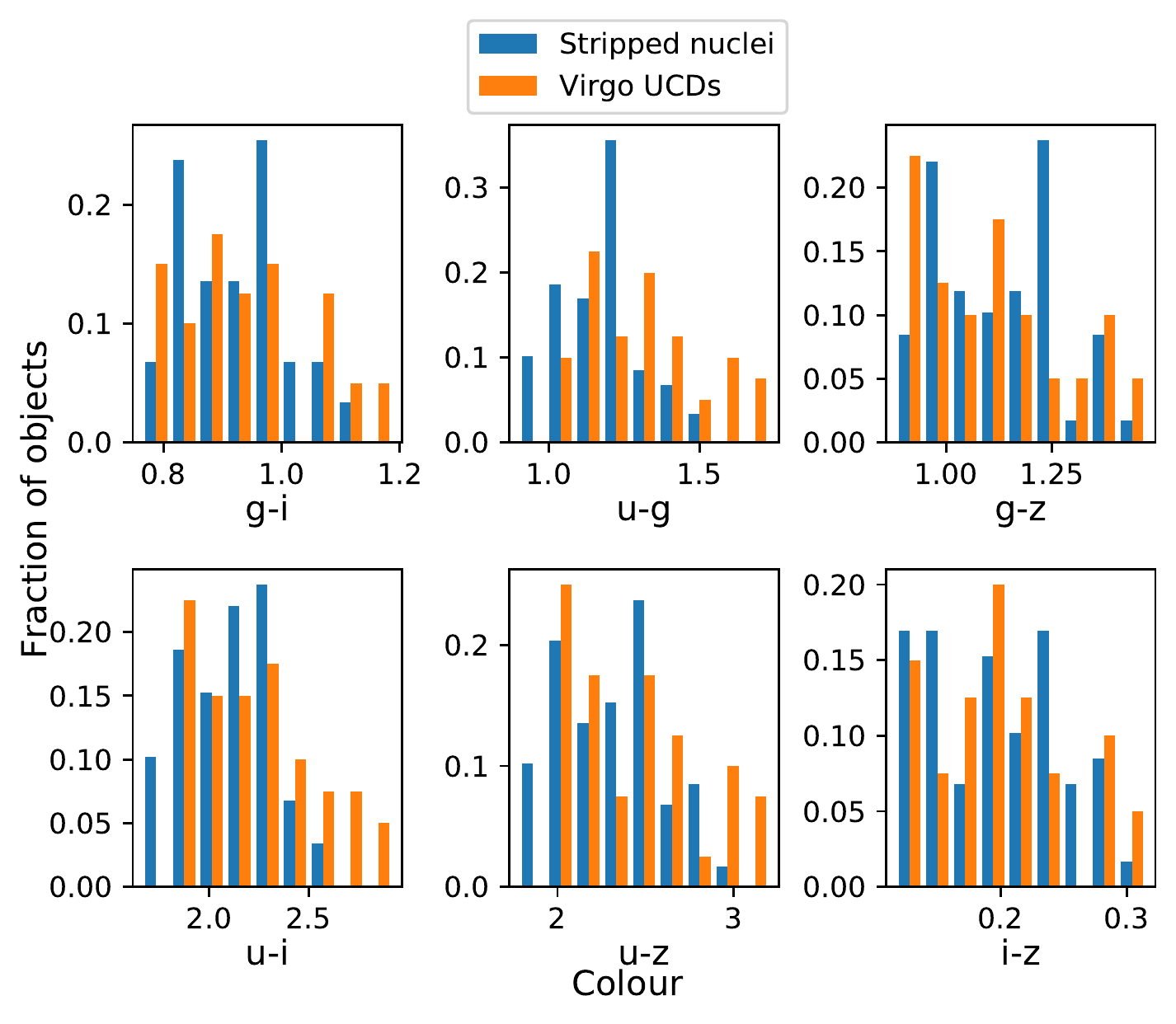}
    \caption{Colours of M > 1~$\times$~10\textsuperscript{7}~\(\textup{M}_\odot\) simulated stripped nuclei from our study and observed UCDs from \citep{liu2020generation}.}
    \label{fig:cols1E7}
\end{figure}

Fig.~\ref{fig:cols} plots 6 different colours comparing simulated stripped nuclei to observed UCDs. The observed and simulated objects overlap fairly closely. The simulated stripped nuclei can explain all but the reddest and bluest observed UCDs. The simulated samples overall tend to trend towards redder colours. A KS test found that all colours were inconsistent. Fig.~\ref{fig:cols1E7} plots only high mass M > 1~$\times$~10\textsuperscript{7}~\(\textup{M}_\odot\) objects and finds more consistency with a KS test finding that 4 out of 6 of the colour distributions are consistent to p > 0.05 and 5 out of 6 to p > 0.01. These plots show that the higher mass simulated stripped nuclei sample trends towards bluer colours than the observed UCDs.
It should be noted that accurately calculating colours for the simulated nuclei is difficult, especially at low metallicities because of sensitivities to abundance changes. The u-band is particularly difficult to accurately calculate as it is sensitive to the horizontal branch, alpha abundances and CNO variations \citep[e.g. Fig 1 in][]{Choi2019}. The two M > 1~$\times$~10\textsuperscript{7}~\(\textup{M}_\odot\) plots with p < 0.05, both involved the u band with u - g having p = 3~$\times$~10\textsuperscript{-4}, and u-i having p = 0.035.

\begin{figure}
	\includegraphics[width=\columnwidth]{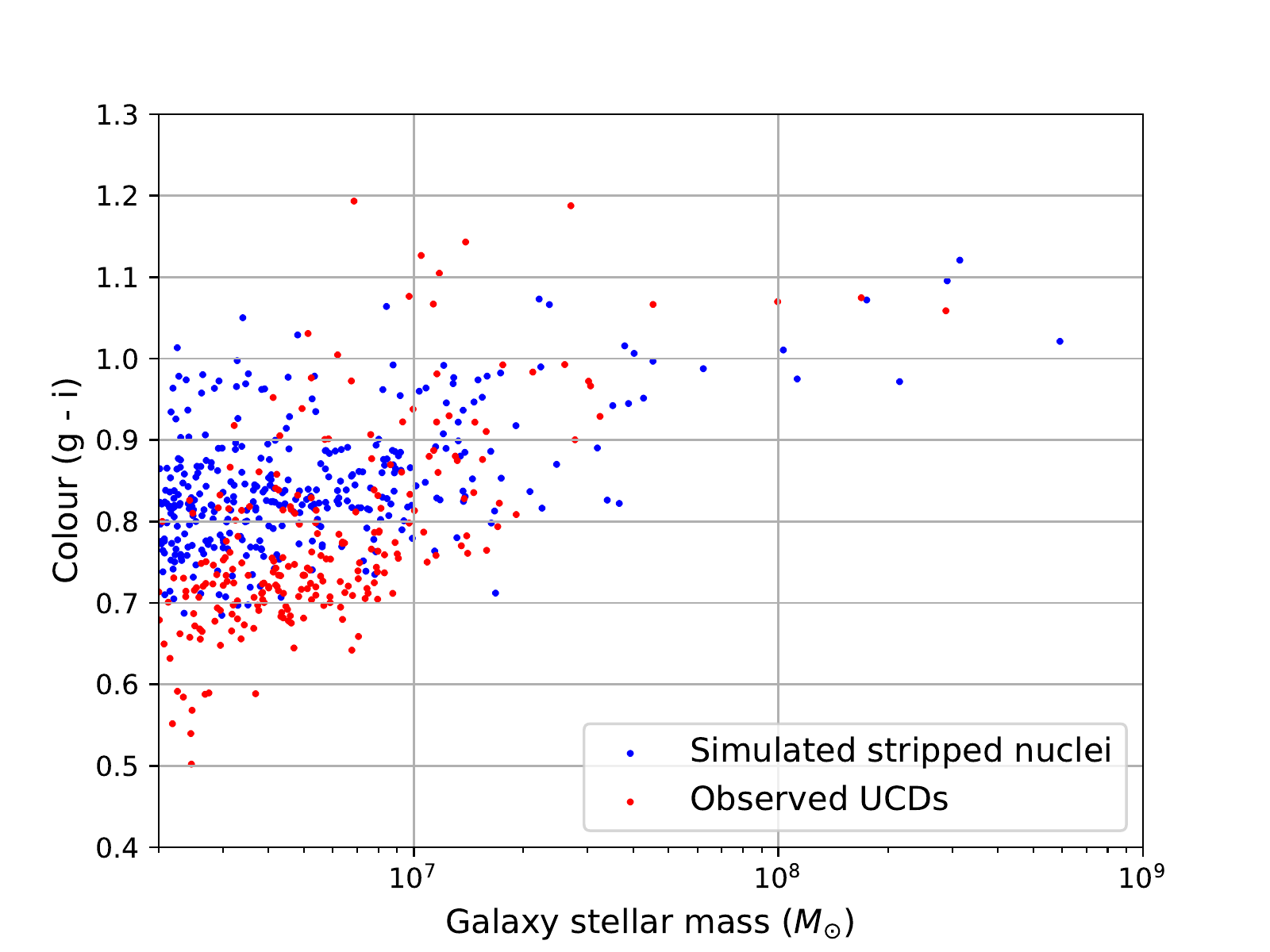}
    \caption{Colour (g-i) against mass for M > 2~$\times$~10\textsuperscript{6}~\(\textup{M}_\odot\) simulated stripped nuclei and observed UCDs from \citep{liu2020generation}.}
    \label{fig:colmass}
\end{figure}

Fig.~\ref{fig:colmass} plots g-i colour against mass and confirms that UCDs and stripped nuclei have a similar colour-mass relationship, with more massive objects being redder, and also that the higher mass objects are more consistent, with the low mass sample being slightly offset from each other.

\begin{figure}
	\includegraphics[width=\columnwidth]{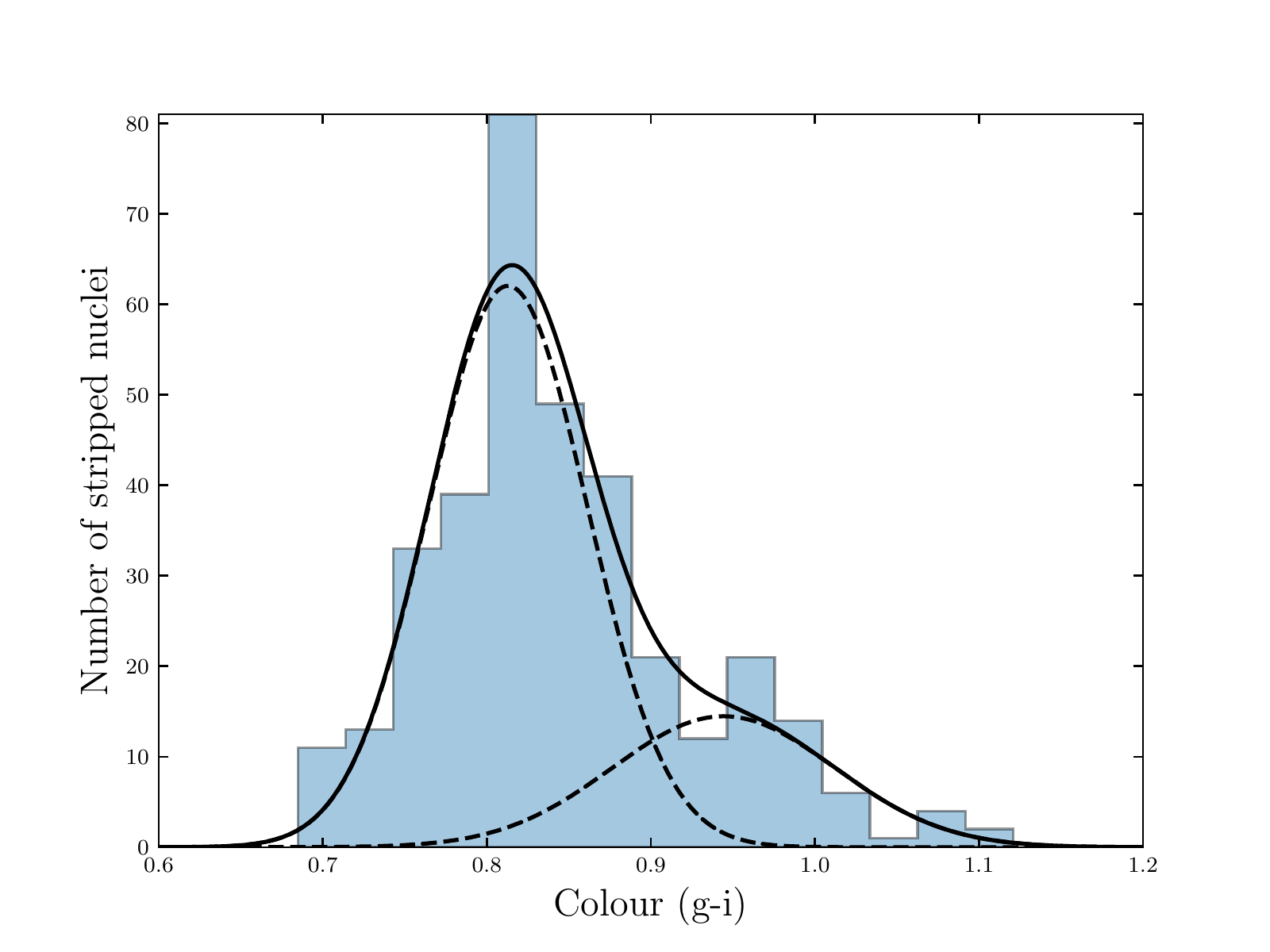}
    \caption{Colour (g-i) bimodality of M > 2~$\times$~10\textsuperscript{6}~\(\textup{M}_\odot\) simulated stripped nuclei from the most massive cluster in the EAGLE simulation.} 
    \label{fig:gicolsunad}
\end{figure}

We find that there is some evidence for a bimodal colour distribution of our sample of simulated stripped nuclei. In Fig.~\ref{fig:gicolsunad} the g-i colour of our sample of simulated stripped nuclei is fitted with a one-dimensional Gaussian mixture model. The model selection criteria Bayesian information criterion (BIC) is minimized for a two component model.
This shows that the colour distribution can appear bimodal, even when the sample is composed of only a single type of objects that do not have a metallicity bimodality.

\begin{figure}
	\includegraphics[width=\columnwidth]{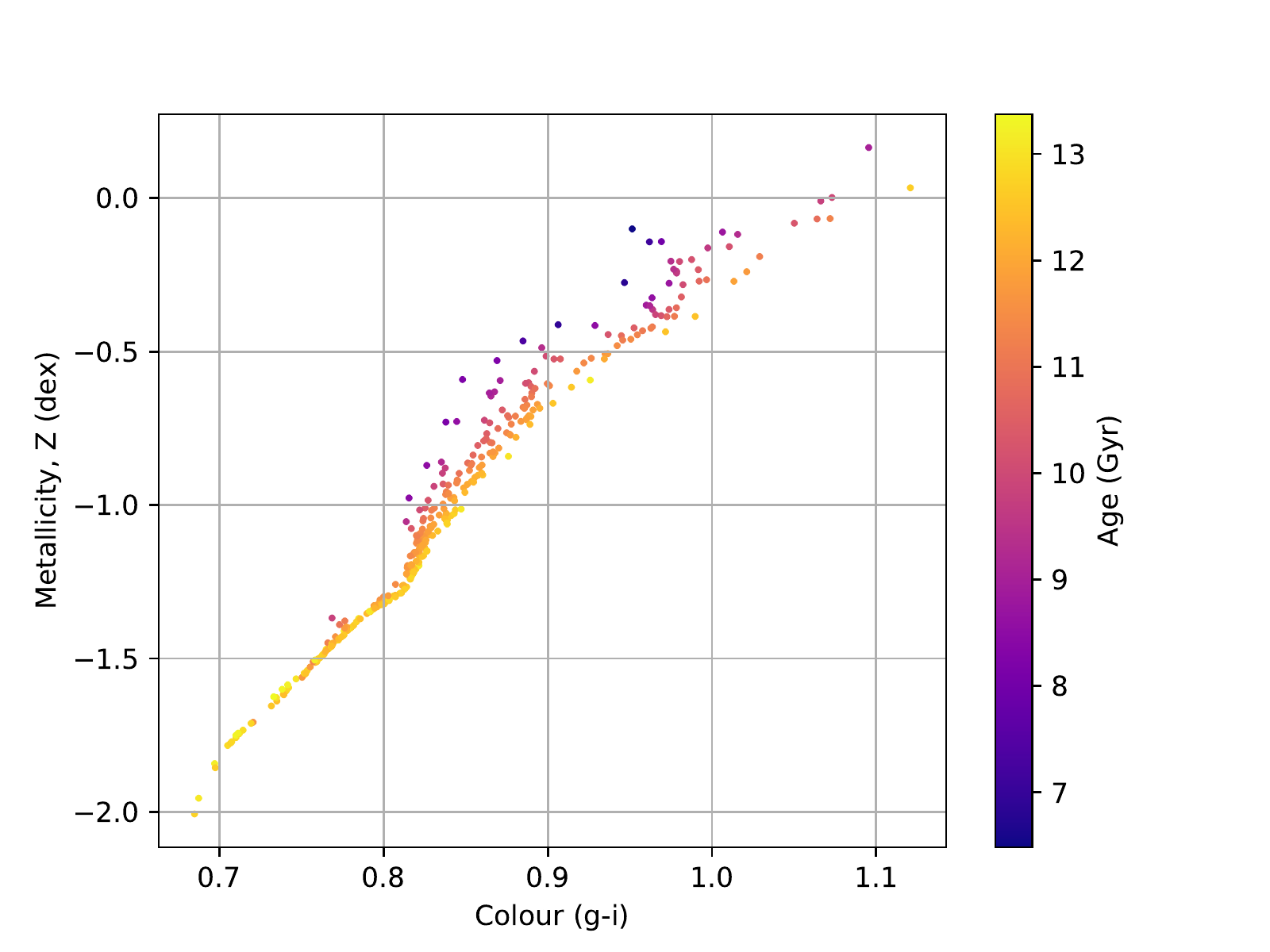}
    \caption{Plot of metallicity against g-i colour-coded by age of M > 2~$\times$~10\textsuperscript{6}~\(\textup{M}_\odot\) stripped nuclei from the most massive simulated cluster in EAGLE}
    \label{fig:agemetcol}
\end{figure}

Fig.~\ref{fig:agemetcol} plots colour against metallicity for the simulated stripped nuclei, with points coloured by stellar formation time age. The plot has a wedge-shaped colour distribution, with inflection points at g-i$\approx$0.8, and g-i$\approx${0.95}. One explanation for the colour bimodality in Fig.~\ref{fig:gicolsunad} is that it stems from the relationship between colour and mass, plotted in Fig.~\ref{fig:colmass} where more massive objects are redder. As the mass increases, the number of objects decreases, but they also trend to redder colours, creating a secondary peak at the redder end of the distribution. Combined with the inflection points in Fig.~\ref{fig:agemetcol} this can create the observed bimodality.

\subsection{Low mass objects with \texorpdfstring{M $<$ 1~$\times$~10\textsuperscript{6}~\(\textup{M}_\odot\)}{Lg}}
\label{section:lowmass}
In \citet{Mayes2020} we found that there should exist a number of stripped nuclei in the mass range of globular clusters. 

\begin{figure}
	\includegraphics[width=\columnwidth]{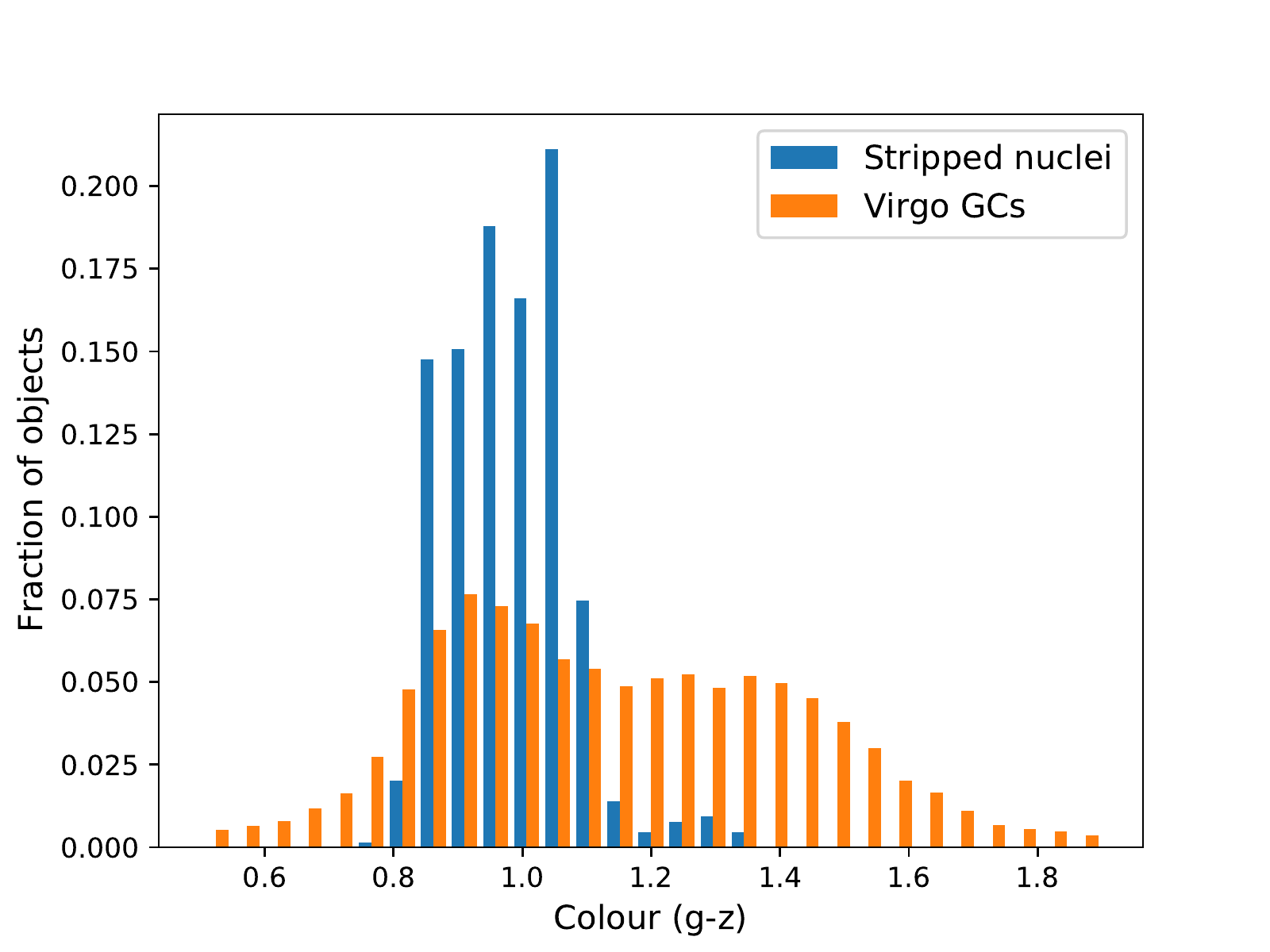}
    \caption{Plot of g-z colours for M < 1~$\times$~10\textsuperscript{6}~\(\textup{M}_\odot\) simulated stripped nuclei from the most massive simulated cluster in EAGLE and Virgo globular clusters from the ACS Virgo survey \citep{jordan2009}}
    \label{fig:gimetage}
\end{figure}

Fig.~\ref{fig:gimetage} plots the distribution of g-z colours for globular clusters belonging to 100 Virgo galaxies observed by the ACS Virgo Cluster survey  (ACSVCS) \citet{jordan2009}, clearly showing the bimodal blue and red populations. 

Note that while the numbers of blue and red globular clusters are approximately equal on this plot, the spatial distribution of blue GCs is less concentrated than that of red GCs, and data from \citet{jordan2009} does not extend far enough from the centers of the ACSVCS galaxies to fully sample the GC distributions at large galactocentric distances. Indeed \citet{Durrell2014}, in a study encompassing the full extent of Virgo, found that there are approximately twice as many blue globular clusters as red ones in the Virgo cluster. This should not impact our results unless the additional blue GCs affect the location of the blue peak in Fig.~\ref{fig:gimetage}.

On this plot we include the population of low mass M < 1~$\times$~10\textsuperscript{6}~\(\textup{M}_\odot\) simulated stripped nuclei. The simulated stripped nuclei appears to strongly overlap with the blue population of globular clusters, with a smaller number overlapping with the red globular clusters. In \citet{Mayes2020} we determined that approximately 1-2 per cent of a massive galaxy’s globular cluster population could be stripped nuclei, here we find that that population will be predominately blue.

\section{Discussion}

\subsection{Can the ages and  metallicities of simulated stripped nuclei explain those of observed UCDs?}

\subsubsection{Ages}
\label{section:agesdisc}
In Section \ref{section:ages} we compared the ages of observed UCDs against the stellar formation times and merger times of simulated stripped nuclei. We found that both UCDs and stripped nuclei are predominantly old objects, with mean ages in all cases being greater than 9 Gyr. The nuclei of progenitor galaxies form approximately 2.5 Gyrs before their host galaxies merge and produce stripped nuclei, with the mean star formation age being 11.47~$\pm$~0.09 Gyr, while the mean merger time is 9.0~$\pm$~0.2 Gyr.

Our sample of observed UCDs in Table~\ref{tab:agesmets} appears slightly younger than the population of simulated stripped nuclei, with a mean age of 10.6 $\pm$ 0.4 Gyr. 
An extended star formation history of some UCDs has been cited as a reason why these UCDs may be stripped nuclei as this distinguishes them from globular clusters \citep{Norris2015, Neumayer_2020}. However the youngest stripped nuclei that the stellar formation time can explain is 5 Gyr, inconsistent with these objects. These young objects could be explained if some stripped nuclei experience a significant starburst at merger that is not accurately measured in EAGLE due to the low particle resolution.
Studies that have compared dwarf galaxy nuclei to UCDs have found that UCDs are typically older than nuclei \citep[e.g.][]{Evstig2007a}. However we find that comparing UCDs to present day dwarf galaxy nuclei is not a fair comparison as the majority of mergers occur >8 Gyr ago (z > 1). Ideally, comparisons between nuclei and UCDs should be made with the z > 1 dwarf galaxy population. 

We conclude that our simulated stripped nuclei are old objects with the mean ages being greater than 9 Gyr, which is consistent with the observations that UCDs are also predominantly old objects. Some young UCDs may have experienced a starburst at the time of merger \citep[e.g.][]{Mihos1994, Barton2000}.

\subsubsection{Metallicity}
In Section \ref{section:midmets} we compared the metallicities, [Fe/H], of observed UCDs with the [Fe/H] of simulated stripped nuclei. We found that the metallicity of our simulated objects could explain the metallicities of observed UCDs within 0.1-0.2 dex, a difference within the uncertainties of the correction applied to the EAGLE metallicities. We predict from our sample of simulated stripped nuclei that there should exist a large population of low metallicity stripped nuclei that have not been observed as UCDs. This could be due to the limited measurements made of UCD metallicities at lower masses, or these stripped nuclei may have been observed as blue globular clusters. 

The metallicities of simulated stripped nuclei were found to be consistent with UCDs above a mass of 1~$\times$~10\textsuperscript{7}~\(\textup{M}_\odot\), with a KS test returning p = 0.22. Both observed UCDs and simulated stripped nuclei appear to have a mass-metallicity relationship, with more massive objects being more metal rich. A UCD mass-metallicity relationship that flattens and connects with compact ellipticals at M > 1~$\times$~10\textsuperscript{8}~\(\textup{M}_\odot\) has been previously reported in the literature \citep[e.g.][]{Zhang2018}, and explained as a result of UCDs possibly being stripped nuclei. Virgo dwarf elliptical nuclei largely follow the UCD mass-metallicity relation, while globular clusters are offset towards higher metallicities at a given mass.

The metallicities of our stripped nuclei sample do not factor in the possibility that some of our stripped nuclei may experience a starburst that the simulation does not account for. In Section \ref{section:agesdisc} we suggested that the lower ages of some observed UCDs could be explained by some stripped nuclei experiencing a starburst at the time of merger. If this is the case these stripped nuclei would also become more metal rich. However only 4 per cent of observed UCDs are too young to be explained by the ages of the stripped nuclei, so only a few stripped nuclei would need to experience starbursts to explain the observed UCDs. Assuming an extreme scenario where 20 per cent of stripped nuclei with ages 12-13 Gyr experience starbursts to give them the same ages and metallicities as those with ages 6-7 Gyr, the metallicity of the sample would increase by $\approx$0.24 dex. This would bring the metallicities of the simulated sample slightly closer to the observed sample. 

The near-solar metallicity of some UCDs has been cited as a reason why they may be stripped nuclei e.g. M59-UCD3 \citep{LiuPeng2015}, VUCD3 and M59cO \citep{Ahn2017}. UCDs sit above the metallicity–luminosity relation of early-type galaxies and have a similar metallicity to dwarf galaxies \citep{Chilingarian2011, Francis2012}. At a mass of 4~$\times$~10\textsuperscript{7}~\(\textup{M}_\odot\) \citet{Janz_2015} observe a transition to extremely high metallicities, only paralleled by nuclei. This is roughly the mass that is believed to be the upper mass limit for the true old star cluster population \citep[e.g.][]{Norris2019} which is consistent with our finding that at M > 1~$\times$~10\textsuperscript{7}~\(\textup{M}_\odot\) simulated stripped nuclei have near-solar metallicities, consistent with observed UCDs. 

It should be noted that UCDs often have lower metallicities and older ages than nuclei \citep{Evstig2007a}, however, the older nuclei that represent UCD progenitors would have been less metal rich than nuclei existing at the present day \citep{Chilingarian2008}. In high-density environments which are conductive to forming stripped nuclei, UCDs and dwarf elliptical nuclei are found to have consistent ages and metallicities \citep{Paudel2010}. Our results are consistent with this, with our high mass stripped nuclei also being less metal rich than present day galaxy nuclei.

\citet{Pfeffer2016} predicted a mass-metallicity relationship for the stripped nuclei much shallower than the mass-metallicity of M87 UCDs observed by \citet{Zhang2018}. \citet{Zhang2018} noted however that \citet{Pfeffer2016}'s stripped nuclei metallicities relied on assigning their stripped nuclei sample a fixed metallicity which is simply offset by a constant amount from that of their host galaxy. This is an issue because the metallicity offset between nuclei and host galaxies is mass dependent, with nuclei of massive galaxies being more metal rich than the host, and nuclei of dwarf galaxies being less metal rich \citep{Neumayer_2020}. 

Our method of using the 5 innermost particles of a galaxy is more self-consistent. Fig.~\ref{fig:nsc} shows the mass comparison between the innermost particles of EAGLE galaxies and their hosts and finds a similar relationship to that plotted in figure 9b of \citet{Neumayer_2020} for higher masses. At low masses our relationship breaks down as the low resolution of the EAGLE simulation means there is no difference in the metallicities of the innermost particles and the host galaxy, so at low masses our simulated stripped nuclei sample may be slightly more metal rich than observed nuclei.

We find that our sample of stripped nuclei in Fig.~\ref{fig:femassplot} have a mass-metallicity relationship that is steeper than that predicted by \citet{Pfeffer2016}, but still somewhat shallower than \citet{Zhang2018}. Fig.~\ref{fig:nsc} appears to have a somewhat shallower relation than figure 9b in \citet{Neumayer_2020}, which may have produced this.

We conclude that the metallicities of UCDs with  M > 1~$\times$~10\textsuperscript{7}~\(\textup{M}_\odot\) can be completely explained by them being stripped nuclei. We predict that there should exist a larger population of low mass, low metallicity stripped nuclei that are not currently observed as UCDs, but may be observed as part of the globular cluster population.

\subsection{Can the colours of simulated stripped nuclei explain those of observed UCDs?}

\subsubsection{Colours of stripped nuclei and UCDs}
A tight colour-magnitude relation exists for nuclei \citep{Hilker_2003}. 
UCDs have shown evidence of a similar colour-magnitude relation \citep{Cote2006,Evstig2008A, brodie2011} that has been considered evidence for them being stripped nuclei. in Fig.~\ref{fig:colmass} we confirm that there exists a colour-mass relationship for simulated stripped nuclei, that follows from the mass-metallicity relationship where more massive and hence brighter simulated stripped nuclei are redder, and the colours of higher mass simulated stripped nuclei are consistent with observed UCDs. 

In Fig.~\ref{fig:cols} we compare the colours of simulated stripped nuclei against observed UCDs from \citet{liu2020generation} for objects with M~>~2~$\times$~10\textsuperscript{6}~\(\textup{M}_\odot\). Visually the colour distributions appear similar, however KS tests found the samples to be inconsistent. This inconsistency could possibly be due to the incompleteness of the low mass UCD sample, or due to uncertainties in the colour calculation of low metallicity stripped nuclei.

Unlike the age and metallicity data, which are compiled from a variety of different papers, the sample from \citet{liu2020generation} is a single uniform sample, although biased towards larger brighter objects.
As a consequence of the observed UCDs being selected to have radii r\textsubscript{h} > 11~pc, the observed UCDs are undersampled at lower masses (below M~$\approx$~10\textsuperscript{7}~\(\textup{M}_\odot\)). Because of the colour-magnitude relationship of UCDs this means that the observed UCD sample may skew slightly redder than a complete sample. However for the lower mass objects the observed UCDs are found to be slightly bluer than the simulated stripped nuclei. This may be a consequence of the difficulty of accurately calculating colour at lower metallicities for our simulated stripped nuclei sample, making colour comparisons at low masses less accurate.

In Fig.~\ref{fig:cols1E7} we compare the colours of simulated stripped nuclei against observed UCDs from \citet{liu2020generation} for objects with M > 1~$\times$~10\textsuperscript{7}~\(\textup{M}_\odot\).  Visually for both the low and high mass plots the colour distributions appear similar, and four out of six of the colour plots are found to be consistent. The two that are found to be inconsistent are those with colours that required the u-band, which is most prone to error due to uncertainties in the bandpass definition and/or changes in the computed spectral energy distribution. 
A starburst at merger that produces metals would also redden the sample of stripped nuclei, however as the predicted metallicity shift is very small this would not affect our results.
The more limited data for low mass UCDs, as well as the difficulty in accurately calculating colour at low metallicities means that we cannot draw any solid conclusions about low mass M < 1~$\times$~10\textsuperscript{7}~\(\textup{M}_\odot\) UCDs. However, we find that the colours of high mass UCDs are consistent with them being stripped nuclei.

\subsubsection{Do stripped nuclei exhibit bimodal colour and metallicity distributions?}

Globular cluster systems in massive elliptical galaxies are known to have a prominent colour bimodality \citep[e.g.][]{Gebhardt1999, Kundua, Peng_2006}. In some systems the number of red and blue objects is approximately equal, while in others one colour dominates \citep{Peng_2006}. 

\citet{liu2020generation} found that UCDs have a slight bimodality that is more pronounced than galaxy nuclei but less pronounced than globular clusters.
In Section \ref{section:midcolours} we found similar evidence for a bimodal fit to our simulated stripped nuclei colours, despite them consisting of a homogenous group of objects formed by a single mechanism. We found no evidence of a metallicity bimodality.

\citet{Lee_2020} have suggested that the colour bimodality in some old globular cluster populations could stem from the nonlinear relationship between colour and metallicity, rather than a metallicity bimodality. Thus the bimodality \citet{liu2020generation} observe in the UCD population could be explained entirely by them being old predominantly metal-poor stripped nuclei.

\subsection{Are low mass stripped nuclei globular clusters?}

An alternate theory for the blue-red globular cluster bimodality, other than it arising from the relationship between colour metallicity and age is that it arises because globular cluster systems of massive galaxies are composite, consisting of both red globular clusters that were formed in-situ by the massive metal rich galaxy, and blue globular clusters that were accreted from smaller metal-poor satellite galaxies \citep[e.g.][]{Cote_1998, Forbes_2010, Leaman_2013}. 

In \citet{Mayes2020} we found that stripped nuclei should contribute a small number of objects ( $\approx$ 1-2 per cent) to the globular cluster populations of massive galaxies and galaxy clusters. Fig.~\ref{fig:gimetage} plots the g-z colours of our sample of low mass M < 1~$\times$~10\textsuperscript{6}~\(\textup{M}_\odot\) simulated stripped nuclei alongside Virgo globular clusters from \citet{jordan2009}. We find that our sample of low mass simulated stripped nuclei overlaps with the blue population of globular clusters which are believed to be composed of the globular cluster populations of merging dwarf galaxies. We therefore conclude that around massive galaxies a small number of blue globular clusters are the stripped nuclei of dwarf galaxies. Possibly these galaxies that became stripped nuclei also contributed their globular cluster systems to the globular cluster populations of those massive galaxies, creating the larger population of blue globular clusters.
\section{Summary}
In this paper, we determine the ages, colours and metallicities of stripped nuclei simulated using the hydrodynamical EAGLE simulations and compare them to observations of UCDs. We find that the stripped nuclei model of UCD formation can explain the properties of high mass (M > 1~$\times$~10\textsuperscript{7}~\(\textup{M}_\odot\)) observed UCDs. Our conclusions are summarised as follows.
\begin{enumerate}
  \item Stripped nuclei are primarily very old objects, that form and merge in the early stages of the universe (t > 9 Gyr).
  \item The old ages of observed UCDs can be explained by them being stripped nuclei. Some young UCDs may have experienced a starburst during the merger event that produced the stripped nucleus.
  \item Both stripped nuclei and UCDs exhibit a mass-metallicity relation, with high mass objects being more metal rich.
  \item Stripped nuclei can completely explain the metallicities of high mass M > 1~$\times$~10\textsuperscript{7}~\(\textup{M}_\odot\) UCDs.
  \item We predict that there should exist a large population of low mass M < 1~$\times$~10\textsuperscript{7}~\(\textup{M}_\odot\), low metallicity stripped nuclei that have not yet been observed as UCDs. These stripped nuclei may have been observed as blue metal-poor globular clusters.
  \item Stripped nuclei can completely explain the colours of high mass M > 1~$\times$~10\textsuperscript{7}~\(\textup{M}_\odot\) UCDs.
  \item The colours of low mass UCDs overlap with the colours of simulated stripped nuclei, but are found not to be consistent with them, possibly due to incompleteness in the low mass sample, or uncertainties in the colour calculation of low metallicity stripped nuclei.
  \item Stripped nuclei have a colour-mass relationship, that can explain the colour-mass relationship of UCDs. In both cases more massive objects are redder. 
  \item Simulated stripped nuclei appear to have a slightly bimodal colour distribution, despite being a homogenous group of objects. This appears to stems from the calculation of colour from age and metallicity. This could explain the observed UCD colour bimodality.
  \item In \citet{Mayes2020} we found several per cent of a massive galaxies globular cluster population are stripped nuclei. In this paper, we show that  these stripped nuclei are predominately blue in colour.
\end{enumerate}
\section*{Acknowledgements}
We thank the referee for the many helpful suggestions that improved this article.

JP gratefully acknowledges financial support from the Australian Research Council's Discovery Projects funding scheme (DP200102574).

C.L. acknowledges support from the National Natural Science Foundation of China (NSFC, Grant No. 11673017, 11833005, 11933003).

This work is based on observations obtained with MegaPrime/MegaCam, a joint project of CFHT and CEA/DAPNIA, at the Canada France Hawaii Telescope (CFHT) which is operated by the National Research Council (NRC) of Canada, the Institut National des Sciences de Univers of the Centre National de la Recherche Scientique (CNRS) of France, and the University of Hawaii. This work is based in part on data products produced at Terapix available at the Canadian Astronomy Data Centre as part of the CanadaFranceHawaii Telescope Legacy Survey, a collaborative project of NRC and CNRS.

This work used the DiRAC@Durham facility managed by the Institute for Computational Cosmology on behalf of the STFC DiRAC HPC Facility (www.dirac.ac.uk). The equipment was funded by BEIS capital funding via STFC capital grants ST/K00042X/1, ST/P002293/1, ST/R002371/1 and ST/S002502/1, Durham University and STFC operations grant ST/R000832/1. DiRAC is part of the National e-Infrastructure.
\section*{Data Availability}

The data underlying this article will be shared on reasonable request to the corresponding author.



\bibliographystyle{mnras}
\bibliography{biblio} 

\begin{thebibliography}{}
\makeatletter
\relax
\def\mn@urlcharsother{\let\do\@makeother \do\$\do\&\do\#\do\^\do\_\do\%\do\~}
\def\mn@doi{\begingroup\mn@urlcharsother \@ifnextchar [ {\mn@doi@}
  {\mn@doi@[]}}
\def\mn@doi@[#1]#2{\def\@tempa{#1}\ifx\@tempa\@empty \href
  {http://dx.doi.org/#2} {doi:#2}\else \href {http://dx.doi.org/#2} {#1}\fi
  \endgroup}
\def\mn@eprint#1#2{\mn@eprint@#1:#2::\@nil}
\def\mn@eprint@arXiv#1{\href {http://arxiv.org/abs/#1} {{\tt arXiv:#1}}}
\def\mn@eprint@dblp#1{\href {http://dblp.uni-trier.de/rec/bibtex/#1.xml}
  {dblp:#1}}
\def\mn@eprint@#1:#2:#3:#4\@nil{\def\@tempa {#1}\def\@tempb {#2}\def\@tempc
  {#3}\ifx \@tempc \@empty \let \@tempc \@tempb \let \@tempb \@tempa \fi \ifx
  \@tempb \@empty \def\@tempb {arXiv}\fi \@ifundefined
  {mn@eprint@\@tempb}{\@tempb:\@tempc}{\expandafter \expandafter \csname
  mn@eprint@\@tempb\endcsname \expandafter{\@tempc}}}

\bibitem[\protect\citeauthoryear{Afanasiev et~al.,}{Afanasiev
  et~al.}{2018}]{Afanasiev_2018}
Afanasiev A.~V.,  et~al., 2018, \mn@doi [\mnras] {10.1093/mnras/sty913}, 477,
  4856–4865

\bibitem[\protect\citeauthoryear{Ahn \& et al.}{Ahn \& et~al.}{2018}]{Ahn2018}
Ahn C.~P.,  et al. 2018, \mn@doi [\apj] {10.3847/1538-4357/aabc57}, \href
  {http://adsabs.harvard.edu/abs/2018ApJ...858..102A} {858, 102}

\bibitem[\protect\citeauthoryear{Ahn et~al.,}{Ahn et~al.}{2017}]{Ahn2017}
Ahn C.~P.,  et~al., 2017, \mn@doi [\apj] {10.3847/1538-4357/aa6972}, 839, 72

\bibitem[\protect\citeauthoryear{{Barton}, {Geller}  \& {Kenyon}}{{Barton}
  et~al.}{2000}]{Barton2000}
{Barton} E.~J.,  {Geller} M.~J.,   {Kenyon} S.~J.,  2000, \mn@doi [\apj]
  {10.1086/308392}, \href
  {https://ui.adsabs.harvard.edu/abs/2000ApJ...530..660B} {530, 660}

\bibitem[\protect\citeauthoryear{{Bassino}, {Muzzio}  \& {Rabolli}}{{Bassino}
  et~al.}{1994}]{Bassino1994}
{Bassino} L.~P.,  {Muzzio} J.~C.,   {Rabolli} M.,  1994, \mn@doi [\apj]
  {10.1086/174514}, \href {http://adsabs.harvard.edu/abs/1994ApJ...431..634B}
  {431, 634}

\bibitem[\protect\citeauthoryear{Bekki \& et al.}{Bekki \&
  et~al.}{2003}]{Bekki2003}
Bekki K.,  et al. 2003, \mn@doi [\mnras] {10.1046/j.1365-8711.2003.06916.x},
  \href {http://adsabs.harvard.edu/abs/2003MNRAS.344..399B} {344, 399}

\bibitem[\protect\citeauthoryear{{Bekki}, {Couch}  \& {Drinkwater}}{{Bekki}
  et~al.}{2001}]{Bekki2001}
{Bekki} K.,  {Couch} W.~J.,   {Drinkwater} M.~J.,  2001, \mn@doi [\apjl]
  {10.1086/320339}, \href {http://adsabs.harvard.edu/abs/2001ApJ...552L.105B}
  {552, L105}

\bibitem[\protect\citeauthoryear{{Binney} \& {Tremaine}}{{Binney} \&
  {Tremaine}}{1987}]{Binney1987}
{Binney} J.,  {Tremaine} S.,  1987, {Galactic dynamics}.
Princeton University Press

\bibitem[\protect\citeauthoryear{Blakeslee \& DeGraaff}{Blakeslee \&
  DeGraaff}{2008}]{Blakeslee_2008}
Blakeslee J.~P.,  DeGraaff R.~B.,  2008, \mn@doi [\aj]
  {10.1088/0004-6256/136/6/2295}, 136, 2295–2305

\bibitem[\protect\citeauthoryear{{Brodie} \& {Strader}}{{Brodie} \&
  {Strader}}{2006}]{Brodie2006}
{Brodie} J.~P.,  {Strader} J.,  2006, \mn@doi [\araa]
  {10.1146/annurev.astro.44.051905.092441}, \href
  {https://ui.adsabs.harvard.edu/abs/2006ARA&A..44..193B} {44, 193}

\bibitem[\protect\citeauthoryear{{Brodie}, {Romanowsky}, {Strader}  \&
  {Forbes}}{{Brodie} et~al.}{2011}]{brodie2011}
{Brodie} J.~P.,  {Romanowsky} A.~J.,  {Strader} J.,   {Forbes} D.~A.,  2011,
  \mn@doi [\aj] {10.1088/0004-6256/142/6/199}, \href
  {https://ui.adsabs.harvard.edu/abs/2011AJ....142..199B} {142, 199}

\bibitem[\protect\citeauthoryear{{Br{\"u}ns} \& {Kroupa}}{{Br{\"u}ns} \&
  {Kroupa}}{2012}]{brun2012}
{Br{\"u}ns} R.~C.,  {Kroupa} P.,  2012, \mn@doi [\aap]
  {10.1051/0004-6361/201219693}, 547, A65

\bibitem[\protect\citeauthoryear{{Br{\"u}ns}, {Kroupa}, {Fellhauer}, {Metz}  \&
  {Assmann}}{{Br{\"u}ns} et~al.}{2011}]{brun2011}
{Br{\"u}ns} R.~C.,  {Kroupa} P.,  {Fellhauer} M.,  {Metz} M.,   {Assmann} P.,
  2011, \mn@doi [\aap] {10.1051/0004-6361/201016220}, 529, A138

\bibitem[\protect\citeauthoryear{{Chabrier}}{{Chabrier}}{2003}]{Chabrier2003}
{Chabrier} G.,  2003, \mn@doi [\pasp] {10.1086/376392}, \href
  {https://ui.adsabs.harvard.edu/abs/2003PASP..115..763C} {115, 763}

\bibitem[\protect\citeauthoryear{{Chilingarian}}{{Chilingarian}}{2009}]{Chilingarian2009}
{Chilingarian} I.~V.,  2009, \mn@doi [\mnras]
  {10.1111/j.1365-2966.2009.14450.x}, \href
  {https://ui.adsabs.harvard.edu/abs/2009MNRAS.394.1229C} {394, 1229}

\bibitem[\protect\citeauthoryear{{Chilingarian} \& {Mamon}}{{Chilingarian} \&
  {Mamon}}{2008}]{Chilingarian2008}
{Chilingarian} I.~V.,  {Mamon} G.~A.,  2008, \mn@doi [\mnras]
  {10.1111/j.1745-3933.2008.00438.x}, \href
  {https://ui.adsabs.harvard.edu/abs/2008MNRAS.385L..83C} {385, L83}

\bibitem[\protect\citeauthoryear{{Chilingarian}, {Mieske}, {Hilker}  \&
  {Infante}}{{Chilingarian} et~al.}{2011}]{Chilingarian2011}
{Chilingarian} I.~V.,  {Mieske} S.,  {Hilker} M.,   {Infante} L.,  2011,
  \mn@doi [\mnras] {10.1111/j.1365-2966.2010.18000.x}, \href
  {https://ui.adsabs.harvard.edu/abs/2011MNRAS.412.1627C} {412, 1627}

\bibitem[\protect\citeauthoryear{{Choi}, {Conroy}  \& {Johnson}}{{Choi}
  et~al.}{2019}]{Choi2019}
{Choi} J.,  {Conroy} C.,   {Johnson} B.~D.,  2019, \mn@doi [\apj]
  {10.3847/1538-4357/aaff67}, \href
  {https://ui.adsabs.harvard.edu/abs/2019ApJ...872..136C} {872, 136}

\bibitem[\protect\citeauthoryear{Conroy}{Conroy}{2013}]{Conroy_2013}
Conroy C.,  2013, \mn@doi [Annual Review of Astronomy and Astrophysics]
  {10.1146/annurev-astro-082812-141017}, 51, 393–455

\bibitem[\protect\citeauthoryear{{Conroy} \& {Gunn}}{{Conroy} \&
  {Gunn}}{2010}]{Conroy2010}
{Conroy} C.,  {Gunn} J.~E.,  2010, \mn@doi [\apj]
  {10.1088/0004-637X/712/2/833}, \href
  {https://ui.adsabs.harvard.edu/abs/2010ApJ...712..833C} {712, 833}

\bibitem[\protect\citeauthoryear{{Conroy}, {Gunn}  \& {White}}{{Conroy}
  et~al.}{2009}]{Conroy2009}
{Conroy} C.,  {Gunn} J.~E.,   {White} M.,  2009, \mn@doi [\apj]
  {10.1088/0004-637X/699/1/486}, \href
  {https://ui.adsabs.harvard.edu/abs/2009ApJ...699..486C} {699, 486}

\bibitem[\protect\citeauthoryear{C{\^o}t{\'e} \& et al.}{C{\^o}t{\'e} \&
  et~al.}{2006}]{Cote2006}
C{\^o}t{\'e} P.,  et al. 2006, \mn@doi [\apjs] {10.1086/504042}, \href
  {http://adsabs.harvard.edu/abs/2006ApJS..165...57C} {165, 57}

\bibitem[\protect\citeauthoryear{Cote, Marzke  \& West}{Cote
  et~al.}{1998}]{Cote_1998}
Cote P.,  Marzke R.~O.,   West M.~J.,  1998, \mn@doi [\apj] {10.1086/305838},
  501, 554–570

\bibitem[\protect\citeauthoryear{{Da Rocha}, {Mieske}, {Georgiev}, {Hilker},
  {Ziegler}  \& {Mendes de Oliveira}}{{Da Rocha} et~al.}{2011}]{DaRocha2011}
{Da Rocha} C.,  {Mieske} S.,  {Georgiev} I.~Y.,  {Hilker} M.,  {Ziegler} B.~L.,
    {Mendes de Oliveira} C.,  2011, \mn@doi [\aap]
  {10.1051/0004-6361/201015353}, 525, A86

\bibitem[\protect\citeauthoryear{De~Bórtoli, Bassino, Caso  \&
  Ennis}{De~Bórtoli et~al.}{2020}]{De_B_rtoli_2020}
De~Bórtoli B.~J.,  Bassino L.~P.,  Caso J.~P.,   Ennis A.~I.,  2020, \mn@doi
  [\mnras] {10.1093/mnras/staa086}, 492, 4313–4324

\bibitem[\protect\citeauthoryear{Drinkwater \& et al.}{Drinkwater \&
  et~al.}{2003}]{Drinkwater2003}
Drinkwater M.~J.,  et al. 2003, \mn@doi [\nat] {10.1038/nature01666}, \href
  {http://adsabs.harvard.edu/abs/2003Natur.423..519D} {423, 519}

\bibitem[\protect\citeauthoryear{{Drinkwater}, {Jones}, {Gregg}  \&
  {Phillipps}}{{Drinkwater} et~al.}{2000}]{Drinkwater2000}
{Drinkwater} M.~J.,  {Jones} J.~B.,  {Gregg} M.~D.,   {Phillipps} S.,  2000,
  \mn@doi [\pasa] {10.1071/AS00034}, \href
  {https://ui.adsabs.harvard.edu/abs/2000PASA...17..227D} {17, 227}

\bibitem[\protect\citeauthoryear{{Drinkwater}, {Gregg}, {Couch}, {Ferguson},
  {Hilker}, {Jones}, {Karick}  \& {Phillipps}}{{Drinkwater}
  et~al.}{2004}]{Drinkwater2004}
{Drinkwater} M.~J.,  {Gregg} M.~D.,  {Couch} W.~J.,  {Ferguson} H.~C.,
  {Hilker} M.,  {Jones} J.~B.,  {Karick} A.,   {Phillipps} S.,  2004, \mn@doi
  [\pasa] {10.1071/AS04048}, \href
  {https://ui.adsabs.harvard.edu/abs/2004PASA...21..375D} {21, 375}

\bibitem[\protect\citeauthoryear{{Durrell} et~al.,}{{Durrell}
  et~al.}{2014}]{Durrell2014}
{Durrell} P.~R.,  et~al., 2014, \mn@doi [\apj] {10.1088/0004-637X/794/2/103},
  \href {https://ui.adsabs.harvard.edu/abs/2014ApJ...794..103D} {794, 103}

\bibitem[\protect\citeauthoryear{{Evstigneeva}, {Drinkwater}, {Jurek}, {Firth},
  {Jones}, {Gregg}  \& {Phillipps}}{{Evstigneeva} et~al.}{2007}]{Evstig2007a}
{Evstigneeva} E.~A.,  {Drinkwater} M.~J.,  {Jurek} R.,  {Firth} P.,  {Jones}
  J.~B.,  {Gregg} M.~D.,   {Phillipps} S.,  2007, \mn@doi [\mnras]
  {10.1111/j.1365-2966.2007.11856.x}, \href
  {https://ui.adsabs.harvard.edu/abs/2007MNRAS.378.1036E} {378, 1036}

\bibitem[\protect\citeauthoryear{Evstigneeva et~al.,}{Evstigneeva
  et~al.}{2008}]{Evstig2008A}
Evstigneeva E.~A.,  et~al., 2008, \mn@doi [\aj] {10.1088/0004-6256/136/1/461},
  136, 461–478

\bibitem[\protect\citeauthoryear{Faifer, Escudero, Scalia, Smith~Castelli,
  Norris, De~Rossi, Forte  \& Cellone}{Faifer et~al.}{2017}]{Faifer_2017}
Faifer F.~R.,  Escudero C.~G.,  Scalia M.~C.,  Smith~Castelli A.~V.,  Norris
  M.,  De~Rossi M.~E.,  Forte J.~C.,   Cellone S.~A.,  2017, \mn@doi [\aap]
  {10.1051/0004-6361/201730493}, 599, L8

\bibitem[\protect\citeauthoryear{{Fellhauer} \& {Kroupa}}{{Fellhauer} \&
  {Kroupa}}{2002}]{Fellhauer2002}
{Fellhauer} M.,  {Kroupa} P.,  2002, \mn@doi [\mnras]
  {10.1046/j.1365-8711.2002.05087.x}, \href
  {https://ui.adsabs.harvard.edu/abs/2002MNRAS.330..642F} {330, 642}

\bibitem[\protect\citeauthoryear{Forbes \& Bridges}{Forbes \&
  Bridges}{2010}]{Forbes_2010}
Forbes D.~A.,  Bridges T.,  2010, \mn@doi [\mnras]
  {10.1111/j.1365-2966.2010.16373.x}

\bibitem[\protect\citeauthoryear{{Forbes}, {Brodie}  \& {Grillmair}}{{Forbes}
  et~al.}{1997}]{Forbes1997}
{Forbes} D.~A.,  {Brodie} J.~P.,   {Grillmair} C.~J.,  1997, \mn@doi [\aj]
  {10.1086/118382}, \href
  {https://ui.adsabs.harvard.edu/abs/1997AJ....113.1652F} {113, 1652}

\bibitem[\protect\citeauthoryear{Forbes, Spitler, Strader, Romanowsky, Brodie
  \& Foster}{Forbes et~al.}{2011}]{Forbes_2011}
Forbes D.~A.,  Spitler L.~R.,  Strader J.,  Romanowsky A.~J.,  Brodie J.~P.,
  Foster C.,  2011, \mn@doi [\mnras] {10.1111/j.1365-2966.2011.18373.x}, 413,
  2943–2949

\bibitem[\protect\citeauthoryear{Forbes, Norris, Strader, Romanowsky, Pota,
  Kannappan, Brodie  \& Huxor}{Forbes et~al.}{2014}]{Forbes_2014}
Forbes D.~A.,  Norris M.~A.,  Strader J.,  Romanowsky A.~J.,  Pota V.,
  Kannappan S.~J.,  Brodie J.~P.,   Huxor A.,  2014, \mn@doi [\mnras]
  {10.1093/mnras/stu1631}, 444, 2993–3003

\bibitem[\protect\citeauthoryear{Forbes, Ferré-Mateu, Durré, Brodie  \&
  Romanowsky}{Forbes et~al.}{2020}]{Forbes_2020}
Forbes D.~A.,  Ferré-Mateu A.,  Durré M.,  Brodie J.~P.,   Romanowsky A.~J.,
  2020, \mn@doi [\mnras] {10.1093/mnras/staa1924}, 497, 765–775

\bibitem[\protect\citeauthoryear{{Francis}, {Drinkwater}, {Chilingarian},
  {Bolt}  \& {Firth}}{{Francis} et~al.}{2012}]{Francis2012}
{Francis} K.~J.,  {Drinkwater} M.~J.,  {Chilingarian} I.~V.,  {Bolt} A.~M.,
  {Firth} P.,  2012, \mn@doi [\mnras] {10.1111/j.1365-2966.2012.21465.x}, \href
  {https://ui.adsabs.harvard.edu/abs/2012MNRAS.425..325F} {425, 325}

\bibitem[\protect\citeauthoryear{{Gallazzi}, {Pasquali}, {Zibetti}  \&
  {Barbera}}{{Gallazzi} et~al.}{2021}]{Gallazzi2021}
{Gallazzi} A.~R.,  {Pasquali} A.,  {Zibetti} S.,   {Barbera} F.~L.,  2021,
  \mn@doi [\mnras] {10.1093/mnras/stab265}, \href
  {https://ui.adsabs.harvard.edu/abs/2021MNRAS.502.4457G} {502, 4457}

\bibitem[\protect\citeauthoryear{{Gebhardt} \& {Kissler-Patig}}{{Gebhardt} \&
  {Kissler-Patig}}{1999}]{Gebhardt1999}
{Gebhardt} K.,  {Kissler-Patig} M.,  1999, \mn@doi [\aj] {10.1086/301059},
  \href {https://ui.adsabs.harvard.edu/abs/1999AJ....118.1526G} {118, 1526}

\bibitem[\protect\citeauthoryear{Georgiev, Böker, Leigh, Lützgendorf  \&
  Neumayer}{Georgiev et~al.}{2016}]{Georgiev_2016}
Georgiev I.~Y.,  Böker T.,  Leigh N.,  Lützgendorf N.,   Neumayer N.,  2016,
  \mn@doi [\mnras] {10.1093/mnras/stw093}, 457, 2122–2138

\bibitem[\protect\citeauthoryear{{Goerdt}, {Moore}, {Kazantzidis}, {Kaufmann},
  {Macci{\`o}}  \& {Stadel}}{{Goerdt} et~al.}{2008}]{Goerdt2008}
{Goerdt} T.,  {Moore} B.,  {Kazantzidis} S.,  {Kaufmann} T.,  {Macci{\`o}}
  A.~V.,   {Stadel} J.,  2008, \mn@doi [\mnras]
  {10.1111/j.1365-2966.2008.12982.x}, \href
  {https://ui.adsabs.harvard.edu/abs/2008MNRAS.385.2136G} {385, 2136}

\bibitem[\protect\citeauthoryear{{Ha{\c s}egan}~{et al.}}{{Ha{\c s}egan}~{et
  al.}}{2005}]{has2005}
{Ha{\c s}egan}~{et al.} M.,  2005, \mn@doi [\apj] {10.1086/430342}, \href
  {http://adsabs.harvard.edu/abs/2005ApJ...627..203H} {627, 203}

\bibitem[\protect\citeauthoryear{{Hau}, {Spitler}, {Forbes}, {Proctor},
  {Strader}, {Mendel}, {Brodie}  \& {Harris}}{{Hau} et~al.}{2009}]{Hau200}
{Hau} G. K.~T.,  {Spitler} L.~R.,  {Forbes} D.~A.,  {Proctor} R.~N.,  {Strader}
  J.,  {Mendel} J.~T.,  {Brodie} J.~P.,   {Harris} W.~E.,  2009, \mn@doi
  [\mnras] {10.1111/j.1745-3933.2009.00618.x}, \href
  {https://ui.adsabs.harvard.edu/abs/2009MNRAS.394L..97H} {394, L97}

\bibitem[\protect\citeauthoryear{{Hilker}, {Infante}, {Vieira}, {Kissler-Patig}
   \& {Richtler}}{{Hilker} et~al.}{1999}]{Hilker1999}
{Hilker} M.,  {Infante} L.,  {Vieira} G.,  {Kissler-Patig} M.,   {Richtler} T.,
   1999, \mn@doi [Astronomy and Astrophysics, Supplement]
  {10.1051/aas:1999434}, 134, 75

\bibitem[\protect\citeauthoryear{Hilker, Mieske  \& Infante}{Hilker
  et~al.}{2003}]{Hilker_2003}
Hilker M.,  Mieske S.,   Infante L.,  2003, \mn@doi [\aap]
  {10.1051/0004-6361:20021766}, 397, L9–L12

\bibitem[\protect\citeauthoryear{{Hilker}, {Mieske}, {Baumgardt}  \&
  {Dabringhausen}}{{Hilker} et~al.}{2008}]{Hilker2008}
{Hilker} M.,  {Mieske} S.,  {Baumgardt} H.,   {Dabringhausen} J.,  2008, in
  {Vesperini} E.,  {Giersz} M.,   {Sills} A.,  eds,  IAU Symposium Vol. 246,
  Dynamical Evolution of Dense Stellar Systems. pp 427--428,
  \mn@doi{10.1017/S1743921308016104}

\bibitem[\protect\citeauthoryear{Janz et~al.,}{Janz et~al.}{2015}]{Janz_2015}
Janz J.,  et~al., 2015, \mn@doi [\mnras] {10.1093/mnras/stv2636}, 456, 617

\bibitem[\protect\citeauthoryear{{Jennings} et~al.,}{{Jennings}
  et~al.}{2015}]{Jennings2015}
{Jennings} Z.~G.,  et~al., 2015, \mn@doi [\apjl] {10.1088/2041-8205/812/1/L10},
  \href {https://ui.adsabs.harvard.edu/abs/2015ApJ...812L..10J} {812, L10}

\bibitem[\protect\citeauthoryear{{Jord{\'a}n} et~al.,}{{Jord{\'a}n}
  et~al.}{2009}]{jordan2009}
{Jord{\'a}n} A.,  et~al., 2009, \mn@doi [\apjs] {10.1088/0067-0049/180/1/54},
  \href {https://ui.adsabs.harvard.edu/abs/2009ApJS..180...54J} {180, 54}

\bibitem[\protect\citeauthoryear{Kirby, Cohen, Guhathakurta, Cheng, Bullock  \&
  Gallazzi}{Kirby et~al.}{2013}]{Kirby_2013}
Kirby E.~N.,  Cohen J.~G.,  Guhathakurta P.,  Cheng L.,  Bullock J.~S.,
  Gallazzi A.,  2013, \mn@doi [\apj] {10.1088/0004-637x/779/2/102}, 779, 102

\bibitem[\protect\citeauthoryear{{Kroupa}}{{Kroupa}}{1998}]{Kroupa1998}
{Kroupa} P.,  1998, \mn@doi [\mnras] {10.1046/j.1365-8711.1998.01892.x}, \href
  {https://ui.adsabs.harvard.edu/abs/1998MNRAS.300..200K} {300, 200}

\bibitem[\protect\citeauthoryear{{Kundu} \& {Whitmore}}{{Kundu} \&
  {Whitmore}}{2001}]{Kundua}
{Kundu} A.,  {Whitmore} B.~C.,  2001, \mn@doi [\aj] {10.1086/321073}, \href
  {https://ui.adsabs.harvard.edu/abs/2001AJ....121.2950K} {121, 2950}

\bibitem[\protect\citeauthoryear{{Lacey} \& {Cole}}{{Lacey} \&
  {Cole}}{1993}]{Lacey1993}
{Lacey} C.,  {Cole} S.,  1993, \mn@doi [\mnras] {10.1093/mnras/262.3.627},
  \href {https://ui.adsabs.harvard.edu/abs/1993MNRAS.262..627L} {262, 627}

\bibitem[\protect\citeauthoryear{Leaman, VandenBerg  \& Mendel}{Leaman
  et~al.}{2013}]{Leaman_2013}
Leaman R.,  VandenBerg D.~A.,   Mendel J.~T.,  2013, \mn@doi [\mnras]
  {10.1093/mnras/stt1540}, 436, 122–135

\bibitem[\protect\citeauthoryear{Lee, Chung  \& Yoon}{Lee
  et~al.}{2020}]{Lee_2020}
Lee S.-Y.,  Chung C.,   Yoon S.-J.,  2020, \mn@doi [\apj]
  {10.3847/1538-4357/abc4e9}, 905, 124

\bibitem[\protect\citeauthoryear{{Liu}~{et al.}}{{Liu}~{et
  al.}}{2015a}]{Liu2015}
{Liu}~{et al.} C.,  2015a, \mn@doi [\apj] {10.1088/0004-637X/812/1/34}, \href
  {http://adsabs.harvard.edu/abs/2015ApJ...812...34L} {812, 34}

\bibitem[\protect\citeauthoryear{{Liu} et~al.,}{{Liu}
  et~al.}{2015b}]{LiuPeng2015}
{Liu} C.,  et~al., 2015b, \mn@doi [\apjl] {10.1088/2041-8205/812/1/L2}, \href
  {https://ui.adsabs.harvard.edu/abs/2015ApJ...812L...2L} {812, L2}

\bibitem[\protect\citeauthoryear{Liu et~al.,}{Liu
  et~al.}{2020}]{liu2020generation}
Liu C.,  et~al., 2020, \apjs, 250

\bibitem[\protect\citeauthoryear{Madrid}{Madrid}{2011}]{Madrid_2011}
Madrid J.~P.,  2011, \mn@doi [\apj] {10.1088/2041-8205/737/1/l13}, 737, L13

\bibitem[\protect\citeauthoryear{Mayes, Drinkwater, Pfeffer, Baumgardt, Liu,
  Ferrarese, Côté  \& Peng}{Mayes et~al.}{2020}]{Mayes2020}
Mayes R.,  Drinkwater M.,  Pfeffer J.,  Baumgardt H.,  Liu C.,  Ferrarese L.,
  Côté P.,   Peng E.,  2020, \mn@doi [\mnras] {10.1093/mnras/staa3731}, 501

\bibitem[\protect\citeauthoryear{{McAlpine} et~al.,}{{McAlpine}
  et~al.}{2016}]{McAlpine2016}
{McAlpine} S.,  et~al., 2016, \mn@doi [Astronomy and Computing]
  {10.1016/j.ascom.2016.02.004}, \href
  {https://ui.adsabs.harvard.edu/abs/2016A&C....15...72M} {15, 72}

\bibitem[\protect\citeauthoryear{Mieske}{Mieske}{2013}]{Mieske2013}
Mieske S. e.~a.,  2013, \mn@doi [\aap] {10.1051/0004-6361/201322167}, 558, A14

\bibitem[\protect\citeauthoryear{{Mieske} \& {Kroupa}}{{Mieske} \&
  {Kroupa}}{2008}]{Mieske2008}
{Mieske} S.,  {Kroupa} P.,  2008, \mn@doi [\apj] {10.1086/528739}, \href
  {http://adsabs.harvard.edu/abs/2008ApJ...677..276M} {677, 276}

\bibitem[\protect\citeauthoryear{{Mieske}, {Hilker}  \& {Infante}}{{Mieske}
  et~al.}{2002}]{Mieske2002}
{Mieske} S.,  {Hilker} M.,   {Infante} L.,  2002, \mn@doi [\aap]
  {10.1051/0004-6361:20011833}, 383, 823

\bibitem[\protect\citeauthoryear{{Mieske}, {Infante}, {Hilker}, {Hertling},
  {Blakeslee}, {Ben{\'\i}tez}, {Ford}  \& {Zekser}}{{Mieske}
  et~al.}{2005}]{Mieske2005}
{Mieske} S.,  {Infante} L.,  {Hilker} M.,  {Hertling} G.,  {Blakeslee} J.~P.,
  {Ben{\'\i}tez} N.,  {Ford} H.,   {Zekser} K.,  2005, \mn@doi [\aap]
  {10.1051/0004-6361:200400119}, \href
  {https://ui.adsabs.harvard.edu/abs/2005A&A...430L..25M} {430, L25}

\bibitem[\protect\citeauthoryear{{Mieske}, {Hilker}, {Infante}  \&
  {Jord{\'a}n}}{{Mieske} et~al.}{2006}]{Mieske2006}
{Mieske} S.,  {Hilker} M.,  {Infante} L.,   {Jord{\'a}n} A.,  2006, \mn@doi
  [\aj] {10.1086/500583}, \href
  {https://ui.adsabs.harvard.edu/abs/2006AJ....131.2442M} {131, 2442}

\bibitem[\protect\citeauthoryear{{Mieske}, {West}  \& {de Oliveira}}{{Mieske}
  et~al.}{2007a}]{Mieske2007b}
{Mieske} S.,  {West} M.~J.,   {de Oliveira} C.~M.,  2007a, in {Saviane} I.,
  {Ivanov} V.~D.,   {Borissova} J.,  eds, Groups of Galaxies in the Nearby
  Universe. p.~103 (\mn@eprint {arXiv} {astro-ph/0603524}),
  \mn@doi{10.1007/978-3-540-71173-5_16}

\bibitem[\protect\citeauthoryear{{Mieske}, {Hilker}, {Jord{\'a}n}, {Infante}
  \& {Kissler-Patig}}{{Mieske} et~al.}{2007b}]{Mieske2007a}
{Mieske} S.,  {Hilker} M.,  {Jord{\'a}n} A.,  {Infante} L.,   {Kissler-Patig}
  M.,  2007b, \mn@doi [\aap] {10.1051/0004-6361:20077631}, 472, 111

\bibitem[\protect\citeauthoryear{Mieske et~al.,}{Mieske
  et~al.}{2008}]{Mieske_2008}
Mieske S.,  et~al., 2008, \mn@doi [\aap] {10.1051/0004-6361:200810077}, 487,
  921

\bibitem[\protect\citeauthoryear{{Mieske}, {Hilker}  \& {Misgeld}}{{Mieske}
  et~al.}{2012}]{Mieske2012}
{Mieske} S.,  {Hilker} M.,   {Misgeld} I.,  2012, \mn@doi [\aap]
  {10.1051/0004-6361/201117634}, 537, A3

\bibitem[\protect\citeauthoryear{{Mihos} \& {Hernquist}}{{Mihos} \&
  {Hernquist}}{1994}]{Mihos1994}
{Mihos} J.~C.,  {Hernquist} L.,  1994, \mn@doi [\apjl] {10.1086/187299}, \href
  {https://ui.adsabs.harvard.edu/abs/1994ApJ...425L..13M} {425, L13}

\bibitem[\protect\citeauthoryear{Mihos et~al.,}{Mihos
  et~al.}{2015}]{Mihos_2015}
Mihos J.~C.,  et~al., 2015, \mn@doi [\apj] {10.1088/2041-8205/809/2/l21}, 809,
  L21

\bibitem[\protect\citeauthoryear{Neumayer, Seth  \& Böker}{Neumayer
  et~al.}{2020}]{Neumayer_2020}
Neumayer N.,  Seth A.,   Böker T.,  2020, \mn@doi [The Astronomy and
  Astrophysics Review] {10.1007/s00159-020-00125-0}, 28

\bibitem[\protect\citeauthoryear{{Norris} \& {Kannappan}}{{Norris} \&
  {Kannappan}}{2011}]{Norris2011}
{Norris} M.~A.,  {Kannappan} S.~J.,  2011, \mn@doi [\mnras]
  {10.1111/j.1365-2966.2011.18440.x}, \href
  {https://ui.adsabs.harvard.edu/abs/2011MNRAS.414..739N} {414, 739}

\bibitem[\protect\citeauthoryear{{Norris}, {Escudero}, {Faifer}, {Kannappan},
  {Forte}  \& {van den Bosch}}{{Norris} et~al.}{2015}]{Norris2015}
{Norris} M.~A.,  {Escudero} C.~G.,  {Faifer} F.~R.,  {Kannappan} S.~J.,
  {Forte} J.~C.,   {van den Bosch} R. C.~E.,  2015, \mn@doi [\mnras]
  {10.1093/mnras/stv1221}, \href
  {https://ui.adsabs.harvard.edu/abs/2015MNRAS.451.3615N} {451, 3615}

\bibitem[\protect\citeauthoryear{{Norris}, {van de Ven}, {Kannappan},
  {Schinnerer}  \& {Leaman}}{{Norris} et~al.}{2019}]{Norris2019}
{Norris} M.~A.,  {van de Ven} G.,  {Kannappan} S.~J.,  {Schinnerer} E.,
  {Leaman} R.,  2019, \mn@doi [\mnras] {10.1093/mnras/stz2096}, \href
  {https://ui.adsabs.harvard.edu/abs/2019MNRAS.488.5400N} {488, 5400}

\bibitem[\protect\citeauthoryear{Paudel, Lisker  \& Janz}{Paudel
  et~al.}{2010}]{Paudel2010}
Paudel S.,  Lisker T.,   Janz J.,  2010, \mn@doi [Astrophysical Journal -
  ASTROPHYS J] {10.1088/2041-8205/724/1/L64}, 724

\bibitem[\protect\citeauthoryear{Peng et~al.,}{Peng et~al.}{2006}]{Peng_2006}
Peng E.~W.,  et~al., 2006, \mn@doi [\apj] {10.1086/498210}, 639, 95–119

\bibitem[\protect\citeauthoryear{Penny, Forbes, Strader, Usher, Brodie  \&
  Romanowsky}{Penny et~al.}{2014}]{Penny_2014}
Penny S.~J.,  Forbes D.~A.,  Strader J.,  Usher C.,  Brodie J.~P.,   Romanowsky
  A.~J.,  2014, \mn@doi [\mnras] {10.1093/mnras/stu232}, 439, 3808

\bibitem[\protect\citeauthoryear{Pfeffer}{Pfeffer}{2014}]{Pfeffer2014}
Pfeffer J. e.~a.,  2014, \mn@doi [\mnras] {10.1093/mnras/stu1705}, \href
  {http://adsabs.harvard.edu/abs/2014MNRAS.444.3670P} {444, 3670}

\bibitem[\protect\citeauthoryear{{Pfeffer} \& {Baumgardt}}{{Pfeffer} \&
  {Baumgardt}}{2013}]{Pfeffer2013}
{Pfeffer} J.,  {Baumgardt} H.,  2013, \mn@doi [\mnras] {10.1093/mnras/stt867},
  \href {http://adsabs.harvard.edu/abs/2013MNRAS.433.1997P} {433, 1997}

\bibitem[\protect\citeauthoryear{Pfeffer, Hilker, Baumgardt  \&
  Griffen}{Pfeffer et~al.}{2016}]{Pfeffer2016}
Pfeffer J.,  Hilker M.,  Baumgardt H.,   Griffen B.~F.,  2016, \mn@doi [\mnras]
  {10.1093/mnras/stw498}, 458, 2492

\bibitem[\protect\citeauthoryear{{Qu} et~al.,}{{Qu} et~al.}{2017}]{Qu2017}
{Qu} Y.,  et~al., 2017, \mn@doi [\mnras] {10.1093/mnras/stw2437}, \href
  {https://ui.adsabs.harvard.edu/abs/2017MNRAS.464.1659Q} {464, 1659}

\bibitem[\protect\citeauthoryear{{S{\'a}nchez-Janssen}
  et~al.,}{{S{\'a}nchez-Janssen} et~al.}{2019}]{Sanchez2019}
{S{\'a}nchez-Janssen} R.,  et~al., 2019, \mn@doi [\apj]
  {10.3847/1538-4357/aaf4fd}, \href
  {https://ui.adsabs.harvard.edu/abs/2019ApJ...878...18S} {878, 18}

\bibitem[\protect\citeauthoryear{Schaye}{Schaye}{2015}]{Schaye2015}
Schaye J. e.~a.,  2015, \mn@doi [\mnras] {10.1093/mnras/stu2058}, \href
  {http://adsabs.harvard.edu/abs/2015MNRAS.446..521S} {446, 521}

\bibitem[\protect\citeauthoryear{Schweizer, Seitzer, Whitmore, Kelson  \&
  Villanueva}{Schweizer et~al.}{2018}]{Schweizer_2018}
Schweizer F.,  Seitzer P.,  Whitmore B.~C.,  Kelson D.~D.,   Villanueva E.~V.,
  2018, \mn@doi [\apj] {10.3847/1538-4357/aaa424}, 853, 54

\bibitem[\protect\citeauthoryear{Seth}{Seth}{2014}]{Seth2014}
Seth A.~C. e.~a.,  2014, \mn@doi [\nat] {10.1038/nature13762}, \href
  {http://adsabs.harvard.edu/abs/2014Natur.513..398S} {513, 398}

\bibitem[\protect\citeauthoryear{Spengler et~al.,}{Spengler
  et~al.}{2017}]{Spengler2017}
Spengler C.,  et~al., 2017, \mn@doi [\apj] {10.3847/1538-4357/aa8a78}, 849

\bibitem[\protect\citeauthoryear{{Voggel}, {Hilker}  \& {Richtler}}{{Voggel}
  et~al.}{2016}]{Voggel2016}
{Voggel} K.,  {Hilker} M.,   {Richtler} T.,  2016, \mn@doi [\aap]
  {10.1051/0004-6361/201527070}, \href
  {https://ui.adsabs.harvard.edu/abs/2016A&A...586A.102V} {586, A102}

\bibitem[\protect\citeauthoryear{Voggel, Seth, Baumgardt, Mieske, Pfeffer  \&
  Rasskazov}{Voggel et~al.}{2019}]{Voggel_2019}
Voggel K.~T.,  Seth A.~C.,  Baumgardt H.,  Mieske S.,  Pfeffer J.,   Rasskazov
  A.,  2019, \mn@doi [\apj] {10.3847/1538-4357/aaf735}, 871, 159

\bibitem[\protect\citeauthoryear{Wehner \& Harris}{Wehner \&
  Harris}{2007}]{Wehner_2007}
Wehner E. M.~H.,  Harris W.~E.,  2007, \mn@doi [\apj] {10.1086/522305}, 668,
  L35–L38

\bibitem[\protect\citeauthoryear{{Zhang} et~al.,}{{Zhang}
  et~al.}{2018}]{Zhang2018}
{Zhang} H.-X.,  et~al., 2018, \mn@doi [\apj] {10.3847/1538-4357/aab88a}, \href
  {https://ui.adsabs.harvard.edu/abs/2018ApJ...858...37Z} {858, 37}

\makeatother
\end{thebibliography}




\appendix
\section{Metallicity correction}
\label{appendix:metcorrec}
As noted in Section \ref{section:agesmets} low mass EAGLE galaxies are systematically more metal rich than observed galaxies \citep{Schaye2015}. To correct the metallicities of EAGLE simulated galaxies we plotted metallicities against galaxy stellar mass for all the galaxies in the most massive simulated cluster at z = 0.1 and then compared these values to those of observed galaxies from \citet{Kirby_2013}. We then carried out a mass based correction by applying fits to the simulated plot, and finding the average difference between the simulated galaxies and observed galaxies at a given mass. We note that the correction we carried out was found to be somewhat larger for higher mass galaxies and smaller for lower mass galaxies than the difference plotted in Figure 13 of \citet{Schaye2015}, which plots log(Z/\(\textup{Z}_\odot\)), while \citet{Kirby_2013} plots [Fe/H].  

\citet{Kirby_2013} finds that observed galaxies have a linear stellar mass-[Fe/H] relation that appears to be roughly continuous for stellar masses of over nine orders of magnitude:

\begin{equation}
<[Fe/H]>~=~(-1.69~\pm~0.03)~+~(0.3~\pm~0.02)~~log(\frac{L_V}{10^6 ~\textup{L}_\odot})
\end{equation}

Plotting the metallicity [Fe/H] of EAGLE galaxies and fitting lines of best fit between different stellar mass ranges we find the relations:

\begin{equation}
[Fe/H]~=~a~+~b~log(M_{*})
\end{equation}

With the stellar mass ranges, a and b shown in Table.~\ref{tab:metcorrec}

\begin{table}
 \caption{EAGLE metallicity correction values}
 \label{tab:metcorrec}
 \begin{tabular}{lcc}
  \hline
  $M_{*}$ & a & b\\
  \hline
  Mass 10\textsuperscript{7.5}~\(\textup{M}_\odot\) to 10\textsuperscript{8}~\(\textup{M}_\odot\) & -3.5 & 0.36\\
  Mass 10\textsuperscript{8}~\(\textup{M}_\odot\) to 10\textsuperscript{8.5}~\(\textup{M}_\odot\) & -2.3 & 0.21\\
  Mass 10\textsuperscript{8.5}~\(\textup{M}_\odot\) to 10\textsuperscript{9}~\(\textup{M}_\odot\) & -3.8 & 0.39\\
  Mass 10\textsuperscript{9}~\(\textup{M}_\odot\) to 10\textsuperscript{9.5}~\(\textup{M}_\odot\) & -3.3 & 0.33\\
  Mass 10\textsuperscript{9.5}~\(\textup{M}_\odot\) to 10\textsuperscript{10}~\(\textup{M}_\odot\) & -2.5 & 0.25\\
  Mass M > 10\textsuperscript{10}~\(\textup{M}_\odot\) & 2 & 0.2\\
  \hline
 \end{tabular}
\end{table}

At a given stripped nuclei progenitor galaxy mass we calculate the expected metallicity using both the observed and simulated equations. The difference between them gives the correction. This correction is then subtracted from the recorded metallicity:

\begin{equation}
[Fe/H]\textsubscript{corrected} = [Fe/H]\textsubscript{current} - ([Fe/H]\textsubscript{Table A1}-[Fe/H]\textsubscript{A1})
\end{equation}

\section{Inner metallicity of EAGLE galaxies}

\begin{figure}
	\includegraphics[width=\columnwidth]{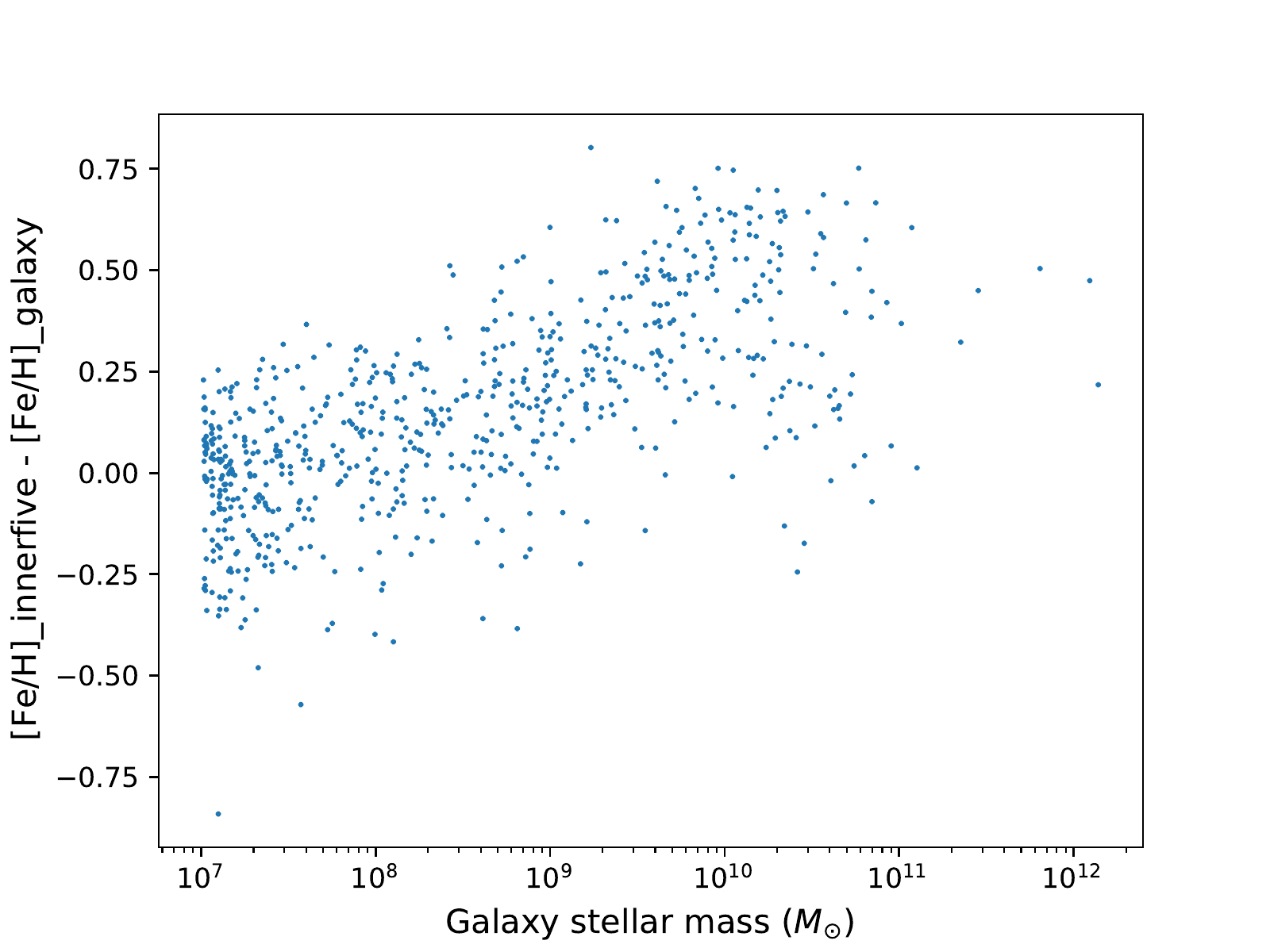}
    \caption{Plot of the difference in metallicities between EAGLE galaxies and their inner five particles}
    \label{fig:nsc}
\end{figure}

Fig.~\ref{fig:nsc} plots the difference in metallicity between EAGLE galaxies and their innermost particles.

\section{Ages and metallicities of observed UCDs}

Table.~\ref{tab:agesmets} shows the metallicities and ages of UCDs compiled from \citet{Mieske_2008,Paudel2010,Francis2012,Chilingarian2011} and \citet{Forbes_2020}. Where studies contained duplicate objects priority was given first to objects with properties measured by spectral fitting and then to more recent papers.

\begin{table*}
	\centering
	\caption{Ages and metallicities of observed UCDs compiled from different sources}
	\label{tab:agesmets}
	\begin{tabular}{llr} 
		\hline
		Object & Age (Gyr) & Metallicity ([Fe/H])\\
		\hline
		F-19 & 12.1 \citep{Chilingarian2011} & -0.19 \citep{Chilingarian2011}\\
		VUCD7 & 10.7 \citep{Paudel2010} & -0.7 \citep{Mieske2008}\\
		VUCD3 & \hl{>15} \citep{Francis2012} & \hl{-0.15} \citep{Francis2012}\\
		UCD1  & \hl{\mbox{13.76 \citep{Francis2012}}} & -\hl{-0.42} \citep{Francis2012}\\
		S417  & - & -0.7 \citep{Mieske2008}\\
		VUCD5  &  \hl{12.24} \citep{Francis2012}  & \hl{-0.57} \citep{Francis2012}\\
		VUCD1  & 11.9 \citep{Paudel2010} & -0.8 \citep{Mieske2008}\\
		F-24  &  >15 \citep{Chilingarian2011}  & -0.67 \citep{Chilingarian2011}\\
		VUCD4  &  11.9 \citep{Paudel2010}  & -1 \citep{Mieske2008}\\
		S999  & 7.6 \citep{Janz_2015} & -1.4 \citep{Mieske2008}\\
		S928  & 7.7 \hl{\mbox{\citep{Forbes_2020}}} & -1.3 \citep{Mieske2008}\\
		UCD5  & \hl{7.21} \citep{Francis2012} & \hl{-1.01} \citep{Francis2012}\\
		VUCD6  & 8.3 \citep{Paudel2010} & -1 \citep{Mieske2008}\\
		F-1  & 13.8 \citep{Chilingarian2011} & -0.64 \citep{Chilingarian2011}\\
		S490  & - & 0.2 \citep{Mieske2008}\\
		F-9  & >15 \citep{Chilingarian2011} & -0.62 \citep{Chilingarian2011}\\
		F-5  & >15 \citep{Chilingarian2011} & -0.34 \citep{Chilingarian2011}\\
		F-6  & 11.1 \citep{Chilingarian2011} & -1.31 \citep{Chilingarian2011}\\
		HCH99-18 & - & -1 \citep{Mieske2008}\\
		F-7  & 14.8 \citep{Chilingarian2011} & -1.2 \citep{Chilingarian2011}\\
		S314  & - & -0.5 \citep{Mieske2008}\\
		F-12  & - & -0.4 \citep{Mieske2008}\\
		G1  & - & -1 \citep{Mieske2008}\\
		HGHH92-C1  & - & -1.2 \citep{Mieske2008}\\
		HGHH92-C23  & - & -1.5 \citep{Mieske2008}\\
		HGHH92-C7  & - & -1.3 \citep{Mieske2008}\\
		F-17  &  >15 \citep{Chilingarian2011}  & \hl{-0.8} \citep{Mieske2008}\\
		F-11  &  >15 \citep{Chilingarian2011}  & -0.61 \citep{Chilingarian2011}\\
		HCH99-15  & - & -1 \citep{Mieske2008}\\
		F-34  &  14.9 \citep{Chilingarian2011}  & -0.77 \citep{Chilingarian2011}\\
		F-22  & >15 \citep{Chilingarian2011} & -0.49 \citep{Chilingarian2011}\\
		HGHH92-C11  & - & -0.5 \citep{Mieske2008}\\
		HGHH92-C17  & - & -1.3 \citep{Mieske2008}\\
		VHH81-C5  & - & -1.6 \citep{Mieske2008}\\
		HGHH92-C21  & - & -1.2 \citep{Mieske2008}\\
		H8005   & - & -1.3 \citep{Mieske2008}\\
		HCH99-2   & - & -1.5 \citep{Mieske2008}\\
		F-53   &  13.8 \citep{Chilingarian2011}  & -0.8 \citep{Chilingarian2011}\\
		HGHH92-C6   & - & -0.9 \citep{Mieske2008}\\
		F-51   & >15 \citep{Chilingarian2011} & -0.23 \citep{Chilingarian2011}\\
		HGHH92-C29   & - & -0.7 \citep{Mieske2008}\\
		HGHH92-C22   & - & -1.2 \citep{Mieske2008}\\
		VHH81-C3   & - & -0.6 \citep{Mieske2008}\\
		HCH99-16   & - & -1.9 \citep{Mieske2008}\\
		HGHH92-C44   & - & -1.6 \citep{Mieske2008}\\
		VUCD2  & \hl{\mbox{9.95 \citep{Francis2012}}} & \hl{-0.98} \citep{Francis2012}\\
		VUCD8  & \hl{\mbox{8.56 \citep{Francis2012}}} & \hl{-1.08} \citep{Francis2012}\\
		VUCD9  & \hl{\mbox{6.81 \citep{Francis2012}}} & \hl{-0.87} \citep{Francis2012}\\
		VUCD10  & 9.8 \citep{Paudel2010} & -\\
		f2  & 12.9 \citep{Chilingarian2011} & -0.73 \citep{Chilingarian2011}\\
		f3  & 13.9 \citep{Chilingarian2011} & -0.61 \citep{Chilingarian2011}\\
		f8  & >15 \citep{Chilingarian2011} & -0.35 \citep{Chilingarian2011}\\
		f13  & 14 \citep{Chilingarian2011} & 0.14 \citep{Chilingarian2011}\\
		f18  & 6.9 \citep{Chilingarian2011} & -0.41 \citep{Chilingarian2011}\\
		f20  & 10.2 \citep{Chilingarian2011} & -1.02 \citep{Chilingarian2011}\\
		f23  & 11.9 \citep{Chilingarian2011} & -0.41 \citep{Chilingarian2011}\\
		f28  & 2 \citep{Chilingarian2011} & -0.94 \citep{Chilingarian2011}\\
		f31  & 10.9 \citep{Chilingarian2011}  & -1.39 \citep{Chilingarian2011}\\
		f46  & 3 \citep{Chilingarian2011} & -0.32 \citep{Chilingarian2011}\\
		f62  & >15 \citep{Chilingarian2011} & -0.26 \citep{Chilingarian2011}\\
		FUCD 0336-3536   & \hl{>15} \citep{Francis2012} & -1.76 \citep{Francis2012}\\
		FUCD 0336-3522   & \hl{>8.18} \citep{Francis2012} & -0.82 \citep{Francis2012}\\
		FUCD 0336-3514   & \hl{>15} \citep{Francis2012} & -0.23 \citep{Francis2012}\\
		FUCD 0336-3548   & \hl{7.98} \citep{Francis2012} & -0.7 \citep{Francis2012}\\
		FUCD 0337-3536   & \hl{7.35} \citep{Francis2012} & -0.87 \citep{Francis2012}\\
		\hline
	\end{tabular}
\end{table*}

\begin{table*}
	\centering
	\label{tab:28000000000000}
	\begin{tabular}{llr} 
		\hline
		Object & Age (Gyr) & Metallicity ([Fe/H])\\
		\hline
		FUCD 0337-3515   & \hl{10.8} \citep{Francis2012} & -0.49 \citep{Francis2012}\\
		FUCD 0338-3513   & \hl{6.24} \citep{Francis2012} & -1.16 \citep{Francis2012}\\
		FUCD 0339-3519   & \hl{13.0} \citep{Francis2012} & -0.46 \citep{Francis2012}\\
		VUCD 1232+0944   & \hl{12.8} \citep{Francis2012} & -1.02 \citep{Francis2012}\\
		VUCD 1233+0952   & \hl{10.9} \citep{Francis2012} & -0.99 \citep{Francis2012}\\
		VUCD 1230+1233   & \hl{9.73} \citep{Francis2012} & -0.76 \citep{Francis2012}\\
		VUCD 1231+1234   & \hl{>15} \citep{Francis2012} & -1.67 \citep{Francis2012}\\
		M60-UCD1   & - & -0.02 \citep{Seth2014}\\
		M59-UCD3   & - & 0 \citep{Ahn2018}\\
		M59cO   & - & 0.1 \citep{Ahn2017}\\
		S8006   & \hl{\mbox{9.15 \citep{Forbes_2020}}} & -\\
		S8005   & \hl{\mbox{9.36 \citep{Forbes_2020}}} & -\\

		\hline
	\end{tabular}
\end{table*}


\bsp	
\label{lastpage}
\end{document}